\documentclass[prb,floatfix,showpacs,twocolumn]{revtex4-2}
\usepackage{amsmath,amssymb,graphicx,color,gensymb}
\usepackage{hyperref}

\def\vecp{\mathbf p}
\def\vecP{\mathbf P}

\def\vecD{\mathbf D}
\def\vS{\boldsymbol \sigma}
\def\vL{\boldsymbol \lambda}
\def\vP{\boldsymbol \Pi}
\newcommand\vecB[1] {\mathbf{#1}}

\def\be{\begin{equation}}
\def\ee{\end{equation}}

\begin{document}
\title{Orientational ordering of water molecules confined in beryl: A theoretical study}

\author{Antonín Klíč}
\thanks{Corresponding author}
\email[e-mail: ]{klic@fzu.cz}
\author{Václav Janiš}
\author{Filip Kadlec}

\affiliation{Institute of Physics, The Czech Academy of Sciences, Na Slovance 2, 18200 Praha  8,  Czech Republic}

\date{\today}

\begin{abstract}
We present an improved model for studying the interactions between dipole moments of water molecules confined in beryl crystals, inspired by recent NMR experiments. Our model is based on a local crystal potential with dihexagonal symmetry for the rotations of water dipole moments, leading to deflection from the $ab$ hexagonal crystallographic plane. This potential shape has significant implications for dipole ordering, which is linked to the non-zero projection of the dipole moment on the hexagonal $c$ axis. To reveal the tendency toward equilibrium-ordered states, we used a variational mean-field approximation, Monte Carlo simulations, and quantum tunneling. Our analysis reveals three types of equilibrium-ordered states: a purely planar dipole order with an antiparallel arrangement in the adjacent planes, a configuration with deflected dipole moments ordered in antiparallel directions, and a helical structure of the dipoles twisting along the $c$ axis.
\end{abstract}

\maketitle

\section{Introduction}

Water molecules confined in various types of environments exhibit properties
substantially different from those of common water phases
\cite{Chakraborty:2017,Chaplin:2010,Thompson:2018,Levinger02,Bampoulis18}. In most of the confined geometries, these
properties are a result of competition between the energy minimization of the
hydrogen-bonded network in water and the constraints due to the confinement, such as
interactions with the cavity surface, possible quantum coherence, or
the necessity to fit within the limited volume. The properties of the usual condensed
phases of water are determined, to a large extent, by the ubiquitous network of
hydrogen bonds. Among the known confined geometries, the opposite extreme case in terms of
hydrogen bonding is represented by crystals containing single molecules of
water, where hydrogen bonding among the water molecules cannot occur because the crystal lattice mutually separates them. Examples of such crystals include gypsum \cite{Bartl72,Yan16,Winkler94,Bishop14}, bassanite \cite{Yan16,Winkler94,Bishop14} and cordierite\cite{Winkler94,Winkler94a,Belyanchikov:2020,Kolesnikov:2014,Dudka2020,Belyanchikov:2020a}, minerals in which low-temperature ordering has been detected. Furthermore,
hydrated beryl ($\rm Be_3Al_2Si_6O_{18}$) is a particular system of
this kind that
came into the focus of researchers more than fifty years ago
\cite{Sugitani1966,Wood67,Gibbs68} and it has been studied till now
\cite{Gorshunov13,Gorshunov14,Zhukova14,Kolesnikov:2014,Kolesnikov:2016,Gorshunov16,Finkelstein17,Belyanchikov:2017,Dressel18,Zhukova:2018,Abalmasov:2021,Belyanchikov:2022,Bassie:2022,Serwatka:2023aa}.
In particular,  the interest in water confined in beryl intensified after
revelations of tendencies of the water dipoles to order anti/ferroelectrically; this ordering develops at low
temperatures with the polarization vectors in the $ab$ plane, i.e.,  perpendicular to he crystallographic $c$ axis \cite{Gorshunov14,Gorshunov16}.

The crystal structure of beryl exhibits hexagonal symmetry (space group
$P6/mcc$), and it contains linear channels of cavities parallel to its hexagonal axis.
Each cavity may host one water molecule. Following the infrared
spectroscopic study of Wood and Nassau \cite{Wood67}, the water molecules have
been assumed to stay in one of two orientations with respect to the
crystal lattice. According to their hypothesis, the molecules occur with the
H--H lines oriented either parallel to the hexagonal $c$ axis (designated as ``type-I water'') or perpendicular to it (``type-II water''). The
occurrence of the latter type was assigned to crystals with additional
doping, as the oxygen atoms of the water molecules may bind to impurity atoms
located within the channels. In the last decade, the incipient ferroelectricity of water
molecules in beryl was demonstrated via spectroscopic measurements of
their collective vibrations in a partially hydrated crystal, revealing a ferroelectric soft phonon mode. This
soft mode was detected in the THz-range spectra of dielectric permittivity, and
its parameters were shown to obey the usual Curie-Weiss, and Cochran temperature
dependence \cite{Gorshunov16}. In that work, a negative Curie
temperature of $T_{\rm C}\approx-20\,K$ was determined from the temperature
dependence of permittivity, implying that the ferroelectric state could not be reached. Notably, at temperatures lower than about 20\,K, the soft phonon vibrational frequency remained almost constant, which was explained
by quantum tunneling of the water molecules between equivalent potential minima \cite{Kolesnikov:2016,Zhukova14}.

\begin{figure*}
\begin{center}
\includegraphics[width=0.8\textwidth]
{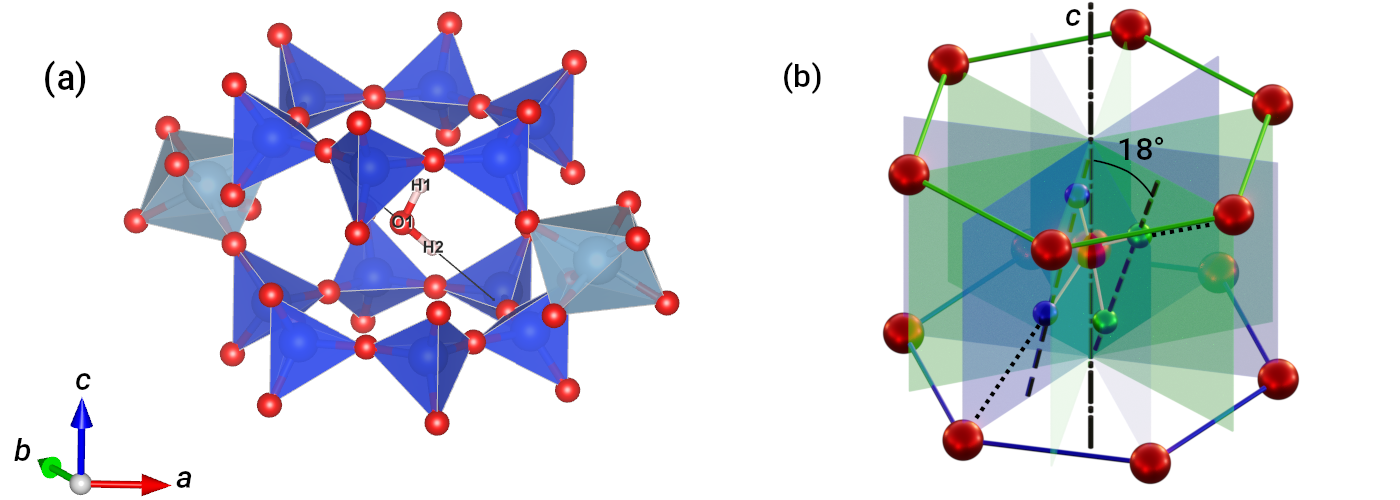}
\end{center}
\caption{(a) Water molecule confined in a void formed by the
  crystal structure of beryl \cite{Gibbs68}. Dark blue: $\rm SiO_4$ tetrahedra,
  pale blue: $\rm AlO_6$ octahedra. The thin black line marks a libration axis
  of the water molecule owing to a transiently formed hydrogen bond between one of the hydrogens and one of the nearest twelve oxygen atoms of the void. The structural data were visualized using the Vesta software \cite{Momma11}. (b)
  Scheme of the void with twelve possible equilibrium orientations of the
  molecule. From the beryl structure, only the twelve O atoms nearest to the
  void center are drawn (red). When the water molecule is constrained to one of the
  planes shown in light blue, an H atom binds to one of the O atoms in the lower
  hexagon (dotted line); then, the H--H line (dashed) intersects the $c$ crystal
axis above the central O atom, and the dipole moment has an upward component
along the $c$ axis. The opposite occurs if the H atoms are constrained to the
planes shown in light green. The same color coding is used in
Fig.~\ref{fig:2x6-Potential}.} \label{fig:12-Structure}
\end{figure*}

Most of the theoretical studies on beryl focused on the
interactions between the dipole moments of the water molecules, and
the models of their collective dynamics assumed
that the type-I molecules can rotate around the hexagonal axis of beryl
so that the H--H lines remain parallel to the hexagonal axis. At the same
time, it was also supposed\cite{Gorshunov13,Kolesnikov:2016,Zhukova14,Belyanchikov:2017,Finkelstein17,Dressel18} that the molecules are subjected, in their angular
orientations to a local potential exhibiting six equivalent minima separated by
an angle of $60\degree$. A recent NMR study of hydrated and deuterated beryl \cite{Chlan22}  provided new data on the orientations and dynamics of the dipoles of water molecules in beryl. These results show that the traditional view of
the molecules' orientations have to be revisited. The numbers and
positions of the NMR lines observed are not compatible with the earlier view
assuming that the H--H lines of the molecules are parallel to the hexagonal axis.
Instead, their mutual angle deduced from these experiments amounts
to about $\pm 18^\circ$ (see Fig.~\ref{fig:12-Structure}).
This suggests that the hydrogen bonds are formed between one of the
hydrogen atoms of water (labeled $\rm H_2$ in
Fig.~\ref{fig:12-Structure}a) and any of the twelve nearest neighboring
oxygen atoms, of which six are in the upper crystal plane, and six are in the lower
one, forming the
cavity. At the same time, the local atomic arrangements of the voids lack
the mirror symmetry with a plane perpendicular to the hexagonal axis;
consequently, the potential minima from the oxygens atoms in the upper and
lower levels surrounding the water molecule are not aligned along the
$c$ axis, but rather alternately angularly tilted (see
Fig.~\ref{fig:12-Structure}b). The observed temperature
dependence of the NMR lines strongly supports a model in which the hydrogen
bonds are present, with the bonded proton of the confined water
molecule forming a rotation axis around which the other free
hydrogen librates near its equilibrium position. It is
known from the structural data \cite{Gibbs68} that the cavity centers are surrounded by
twelve oxygen atoms located at equal distances from the centers. These atoms
are distributed in two layers parallel to the $ab$ planes, namely  at the vertices
of two hexagons (see Fig.~\ref{fig:12-Structure}a). Consequently, each water molecule's possible equilibrium dipole positions within its beryl cavity must equal twelve.

The aim of this paper is to propose a model of the
water molecules in the cavities of the beryl crystal with a confining potential showing dihexagonal symmetry of twelve local minima for the equilibrium positions for their dipole moments. We show that the presence
of twelve energetically equivalent orientations of the water molecules has new and unexpected consequences.  New ordering
  tendencies along the crystallographic $c$ axis emerge, leading to a wider
    variety of dipole states than assumed up to now. In particular, we
      find that the ground state at low temperatures is linked to a chiral ordering along the  $c$ axis. The new experimental findings indicate that the existing models of water
      molecules confined in the cavities of the beryl crystal with a six-fold potential
      degeneracy still need to be completed, revisited, and extended to be compatible with
      the latest observed behavior. Our model is the first step in this direction.

\section{Microscopic model}

\begin{figure}
\begin{center}
\resizebox{0.6\columnwidth}{!}{\includegraphics{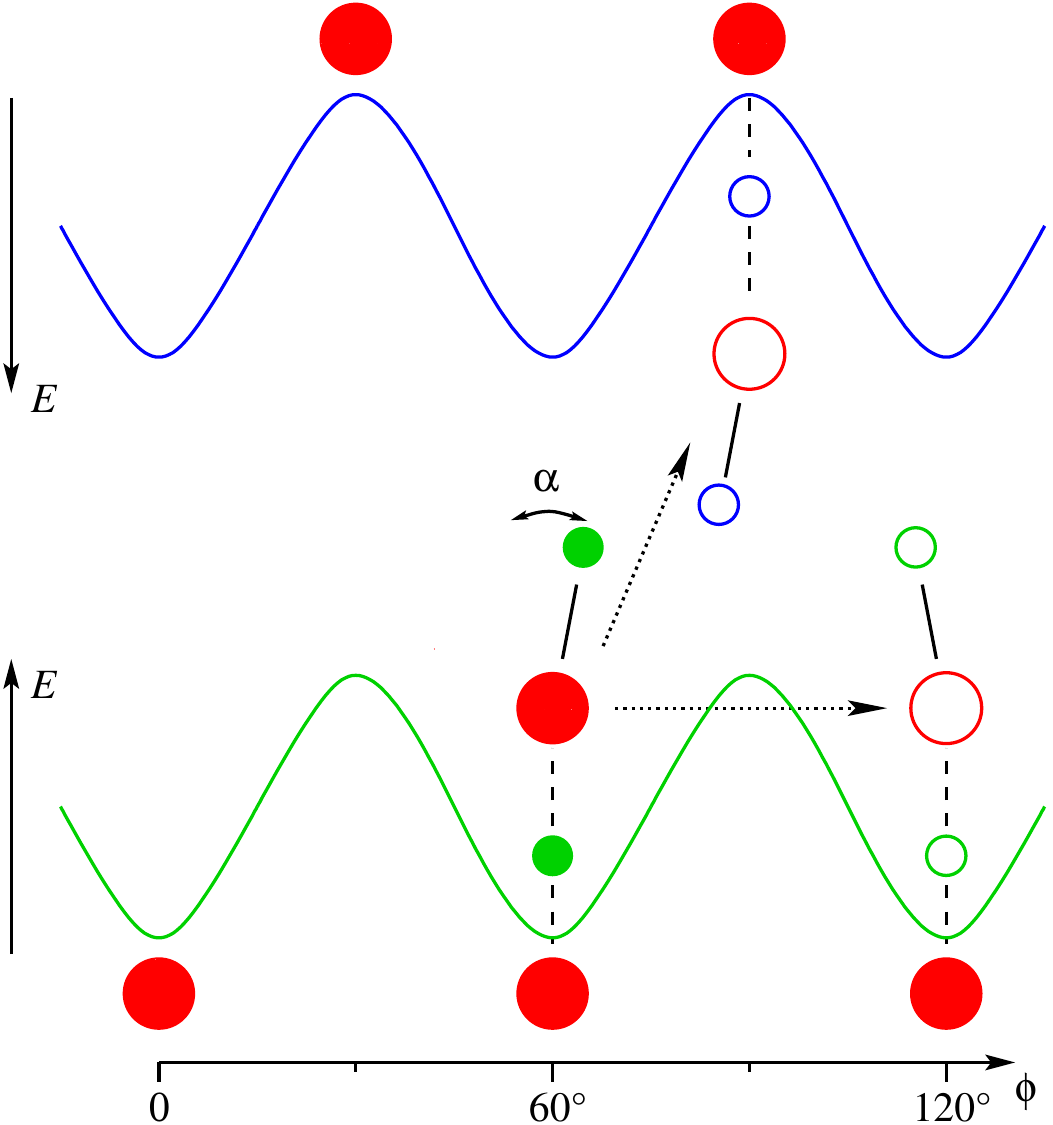}}
\end{center}
\caption{Scheme of angular reorientations of the water molecule in the beryl
cavity. Red circles denote oxygen atoms of both the beryl structure (upper and
lower rows) and the water molecule (central area), whereas small blue and
green circles stand for hydrogen nuclei. These are attracted by the potential
of the cavity (wavy lines; note the opposite signs of the energy scales)
which attracts either of the hydrogen nuclei such that a hydrogen bond is formed
(dashed lines). The water molecule represented by filled circles corresponds to
one orientational configuration of the hydrogen nuclei among the twelve possible
ones. The unbound nucleus is partially free, exhibiting librations (angle
$\alpha$) around its equilibrium position. Reorientations of the molecule
(dotted arrows) may occur in two ways, either via tunneling of the
nucleus through the potential barrier, or when an oxygen
atom from the opposite side of the cavity traps the librating nucleus, breaking
the previous hydrogen bond and freeing the bound nucleus from its previous position. The horizontal axis does not relate to the water molecule's oxygen atom fixed in the cavity center. }
\label{fig:2x6-Potential}
\end{figure}

The water molecules are confined in cavities in the beryl crystal with an
overall hexagonal symmetry along the $c$-axis (also called "vertical direction" in the following).  The
center of mass of the water molecules is positioned at the symmetry axis, and their fluctuations can be neglected. The protons of the hydrogen atoms
in the water molecules are attracted to the oxygens of the lattice ions, forming a nanocavity for the water molecule. Recent NMR experiments
\cite{Chlan22} show that in contrast with the widespread assumption of
previous theoretical works, the H--H lines of the water molecules are tilted
with respect to the $c$ axis. To set up a model potential for the water
molecules' reorientations, see Fig.~\ref{fig:2x6-Potential}; we took advantage of this finding together with the
data on the local crystal arrangement \cite{Gibbs68} showing that six oxygen
atoms are positioned slightly above the center of the cavity and
six below. The two hexagons are not mirrored
by the $ab$ plane but angularly turned by $30\degree$, forming a perfect dihexagonal spatial structure confining the water
molecule in its center of origin. The positions of the oxygen
hexagons, probably forming transient hydrogen bonds to the molecules, are at
the origin of the angular deflection of the dipole moment of water I from the
$ab$ plane. The dipole moment can be directed towards one of the oxygens either
above or below the center of mass of the water molecule. The deflection angle
does not change much with temperature, and its  experimental value
$\theta\approx 18\degree$ \cite{Chlan22} was taken as fixed in the
model description. We hence assume dynamics only in the orbital angle $\phi$
within the $ab$ plane.


The dipole moment of the water molecule in the cavity with only rotational degrees of freedom can be described by a quantum-mechanical Hamiltonian
\begin{equation}\label{eq:H-Loc}
H^{loc} = \frac 1{2I} L^{2} + V^{loc}
\end{equation}
where $L$ is its angular momentum, $I$ its moment of inertia, and
$V^{loc}$ is a local potential due to the surrounding walls. We assume that
angle $\phi$ in the $ab$ plane orthogonal to the symmetry axis fully describes the rotational degrees of freedom. We neglect the dipole dynamics out of the plane since the angle of deflection from the plane is small, and the energy connected with the vertical movements is negligible.

The local potential of the cavity has six minima for each oxygen hexagon. Since the single water molecule is attracted alternately to two hexagons, we can represent the local potential as a $2\times 2$ matrix
\begin{equation}
V^{loc}(\phi) = \frac{V_b}{2}\left(\begin{matrix} 1+\cos 6\phi & \omega/V_b \\ \omega/V_b & 1-\cos 6\phi  \end{matrix}\right) \,,
\end{equation}
where $V_{b}$ is the potential barrier as in the standard six-fold model and
$\omega$ is the amplitude of the transition of the hydrogen bond from the upper
to the lower nearest oxygen. As the lower and upper limits of the potential barrier, we take the values of Ref.~\cite{Kolesnikov:2016},
$V_{b1}=56\,\rm meV$ and $V_{b2}=176\,\rm meV$. We will set the value of $\omega$ selfconsistently (see  Sec.~III.B).

The water dipole moments interact via a binary dipole-dipole interaction, and the full lattice Hamiltonian reads
\begin{subequations}\label{eq:HamiltonianFull}
\begin{multline}
    H= \sum_{i}H^{loc}_i + \frac{1}{4\pi \epsilon_0 \epsilon_r {r_0}^3}\frac{1}{2}\sum_{i\neq j}\sum_{\alpha,\beta}p_i^{\alpha} D_{i j}^{\alpha \beta}p_j^{\beta}
    \\
    - \sum_{i,\alpha}E^{\alpha}_i p_i^{\alpha}
\end{multline}
where $D_{i j}^{\alpha \beta}$ is the dipolar kernel
\begin{equation}
D_{i j}^{\alpha \beta}=\frac{\delta^{\alpha\beta}}{r_{ij}^3}-3\frac{r_{ij}^{\alpha}r_{ij}^{\beta}}{r_{ij}^5}
\end{equation}
\end{subequations}
with dimensionless intermolecular distances $r_{ij}$ (dependent on the lattice structure), $p_i^{\alpha}$ is a dipole vector of molecule $i$, and $\vecB{E}_{i}$ is the electric field at site $i$. For further calculations, we define an interaction constant
$\bar{J}= p_0^2/(4\pi \epsilon_0 \epsilon_rr_0^3)$, where
$\epsilon_r=7$ is the background permittivity \cite{Gorshunov16}, $p_0=1.85\,\rm D$ is the magnitude of dipole
moment of the water molecule, and $r_0=9.2\,\AA$ is the distance between the nearest dipoles in the $ab$ plane (that is $r_{0}/2$ in the $c$ direction).

\section{Single water molecule}

Before approaching the full lattice problem, we resolve the local problem of a single water molecule in the beryl cavity.

\subsection{Ground state energies}

\begin{figure}
\begin{center}
\resizebox{0.9\columnwidth}{!}{\includegraphics{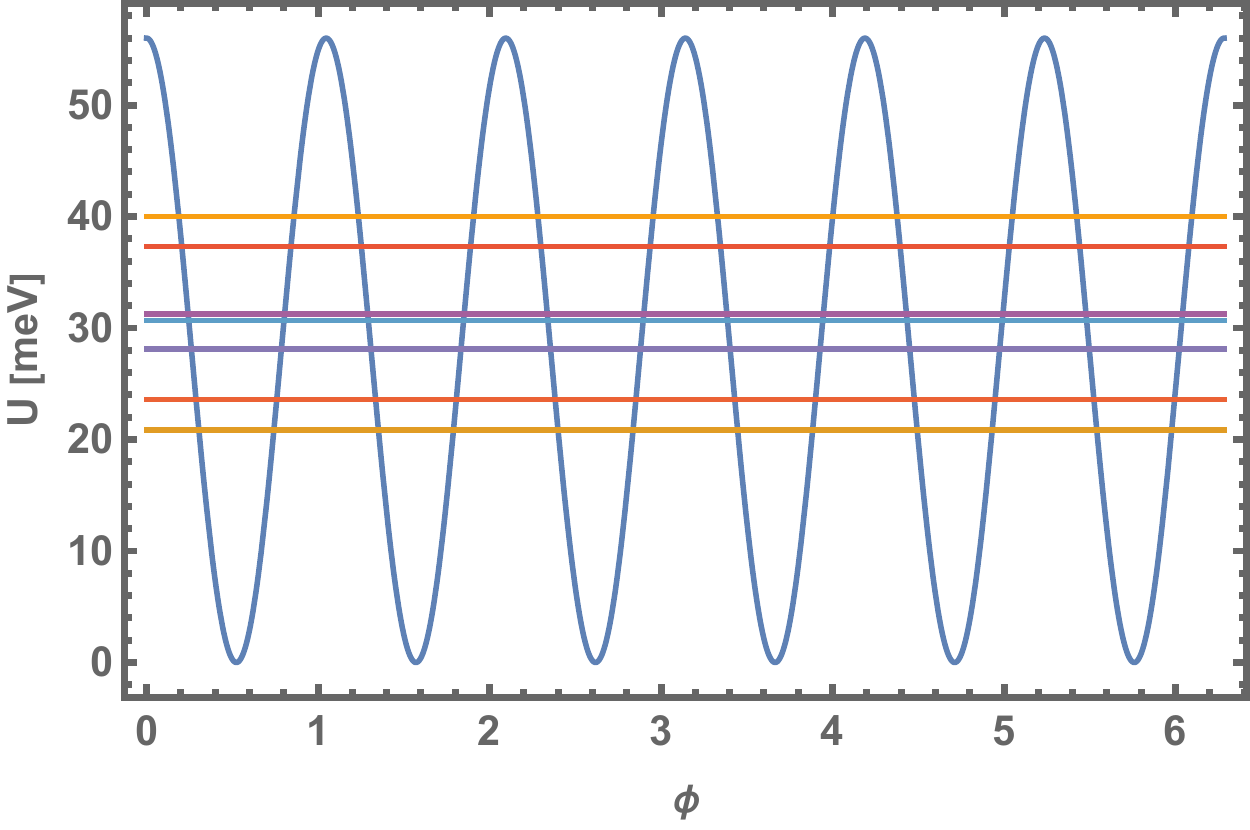}}
\end{center}
\caption{The lowest twelve energy levels of the clock model with the potential
barrier $V_{b1}=56\,\rm meV$ and the transition amplitude between the layers
$\omega=8.2\,\rm meV$.}
\label{figEnergyLevels}
\end{figure}

The cavity potential (see Fig.~\ref{fig:2x6-Potential}) has two six-fold degenerate minima. The eigenvalue problem for the local Hamiltonian $H^{loc}$ from Eq.~\eqref{eq:H-Loc} leads to twelve ground-state energies. They are obtained from the equation
\begin{equation}
    \left(\frac{-\hbar^2}{2 I}+V^{loc}(\phi)\right)\left(\begin{matrix} \psi_{+}(\phi) \\ \psi_{-}(\phi)\end{matrix}\right)
     =E \left(\begin{matrix} \psi_{+}(\phi) \\ \psi_{-}(\phi)\end{matrix}\right)
\end{equation}
with periodic boundary conditions for the planar angle $\phi$.
For the rotational constant $\hbar^2/2I$ of water molecule we choose value $~3.04\,\rm meV$,~\cite{Kolesnikov:2016}.
We numerically solve the eigenvalue equations with a small value $\omega/V_{b}\ll 1$ in order to obtain explicit $\omega$-depdendence of the eigenenergies.
We end up with seven different lowest energy states, five doubly degenerate.
Their explicit values in the interval between  $V_{b1}=56\,\rm meV$ and  $V_{b2}=176\,\rm meV$, with $\omega$-dependence, are
\begin{subequations}\label{eq:eigenEn}
\begin{eqnarray}
E_0&=& 24.5 -26.3 (\omega/V_b) \nonumber  \\
E_{1,2}&=& 27.16 -25.7 (\omega/V_b)\nonumber  \\
E_3&=&24.5 +26.3 (\omega/V_b)\nonumber  \\
E_{4,5}&=& 27.16 +25.7 (\omega/V_b)\nonumber  \\
E_{6,7}&=& 34.4 -22.0 (\omega/V_b) \nonumber \\
E_{8,9}&=& 34.4 +22.0 (\omega/V_b)\nonumber  \\
E_{10,11}&=& 40.5 \nonumber
\end{eqnarray}
\end{subequations}
The eigenenergies depend on $V_b$ and are given in meV units. The boundary values correspond to the extreme values $V_{b1}$ and $V_{b2}$.
These energies determine  the phase space
for a single dipole moment.  The seven quantum energies for $\omega=\omega_1$ are plotted in
Fig.~\ref{figEnergyLevels} together with the local potential
$V^{loc}(\phi)$
with six minima from which we can assess the impact of quantum dynamics.

\subsection{Dynamics: Discrete clock model}

We simplify the dynamics of the dipole within the cavity by explicitly considering only discrete amplitudes of transitions between nearest neighbor local minima, instead of a continuous movement. We introduce three energy parameters: $e_{0}$, $a$, and $d$, which correspond to the ground-state energy and hopping to the locally nearest and next-nearest minima of the six-fold degenerate local potential $V^{loc}$. Note that the parameter $a$ represents the amplitude for reversing the deflection of the dipole moment from the $ab$ plane in the direction of the $c$ axis. The parameter $d$ describes the hopping between two adjacent minima without changing the orientation of the dipole moment along the $c$ axis. The Hamiltonian of the local dipole dynamics reduces the position representation to a matrix
\begin{multline}\label{eq:TwelveMatrix}
\hat{h}^{loc} = e_0 \hat{I}-
\left(
\begin{array}{cccccccccccc}
 0 & a & d & 0 & 0 & 0 & 0 & 0 & 0 & 0 & d & a \\
 a & 0 & a & d & 0 & 0 & 0 & 0 & 0 & 0 & 0 & d \\
 d & a & 0 & a & d & 0 & 0 & 0 & 0 & 0 & 0 & 0 \\
 0 & d & a & 0 & a & d & 0 & 0 & 0 & 0 & 0 & 0 \\
 0 & 0 & d & a & 0 & a & d & 0 & 0 & 0 & 0 & 0 \\
 0 & 0 & 0 & d & a & 0 & a & d & 0 & 0 & 0 & 0 \\
 0 & 0 & 0 & 0 & d & a & 0 & a & d & 0 & 0 & 0 \\
 0 & 0 & 0 & 0 & 0 & d & a & 0 & a & d & 0 & 0 \\
 0 & 0 & 0 & 0 & 0 & 0 & d & a & 0 & a & d & 0 \\
 0 & 0 & 0 & 0 & 0 & 0 & 0 & d & a & 0 & a & d \\
 d & 0 & 0 & 0 & 0 & 0 & 0 & 0 & d & a & 0 & a \\
 a & d & 0 & 0 & 0 & 0 & 0 & 0 & 0 & d & a & 0 \\
\end{array}
\right)
\end{multline}
where $\hat{I}$ is the diagonal matrix. The three model parameters lead to
the following eigenvalues of the above matrix:
$\bar{E}_0=-2 a-2 d+e_0, \bar{E}_3=2 a-2 d+e_0, \bar{E}_{1,2}=\sqrt{3}
a-d+e_0, \bar{E}_{4,5}=\sqrt{3}
a-d+e_0,\bar{E}_{6,7}=-a+d+e_0,\bar{E}_{8,9}=a+d+e_0,\bar{E}_{10,11}=2 d+e_0$.
These eigenenergies should equal
the ground-state energies obtained in the preceding subsection. There are seven independent parameters in matrix \eqref{eq:TwelveMatrix}.
The full set of eigenenergies would make the solution awkward and difficult to control. We use in our simplified matrix only three eigenenergies, namely the lowest $E_{0}$, the highest $E_{10,11}$, and energy $E_{3}$ of the singlet state.  It is important to note that this reduction does not change the result qualitatively. The equations determining $e_{0}, a, d$ then read
 \begin{subequations}
 \begin{eqnarray}
   -2 a-2 d+e_0 &=& E_0 \\
    2 d+e_0 &=& E_{10,11} \\
   2 a-2 d+e_0 &=& E_3 \,.
 \end{eqnarray}
 \end{subequations}

The solution of these equations is $e_0\simeq 32.2\,\rm meV, d= 4\,\rm meV\,-0.002\, \omega^2$ and $a= 0.23\, \omega $. We assume that the probability of jumps between the upper and lower sites is the same as between the left and right sites, which requires choosing $d= 2 a$. From this condition, we obtain the value of $\omega$. As a result, we get these parameter values for the lower bound $V_{b1}=56\,\rm meV$, $\omega_1= 8.2\,\rm meV, d_1= 3.9\,\rm meV, a_1= 1.9\,\rm meV$ and for the upper bound $V_{b2}=176\,\rm meV$, $\omega_2= 3.9\,\rm meV, d_2= 1.26\,\rm meV, a_2=0.63\,\rm meV$.

We use these values to characterize the local parameters of individual dipoles in the description of the water molecules in the extended beryl crystal with nonlocal dipole-dipole interaction from Hamiltonian in Eq.~\eqref{eq:HamiltonianFull}.

\section{Interacting water dipoles in beryl crystal}

\subsection{Ground state}

Having resolved the behavior of a single water molecule in the beryl cavity, we extend the description to the behavior of water molecules in
  the extended beryl crystal. We assume a perfect homogeneity and neglect
  the dilution of water molecules. We describe the model within a tight-binding
  approximation with each unit cell containing one water
  molecule, the local behavior of which will be approximated by the clock model
  with twelve-fold degenerate ground-state energy. We first neglect quantum
  tunneling between the minima.  The rotation angle $\phi$ will describe the positions of the potential minima in the $ab$ plane. Its $n_S=12$ values $\phi_i \in \{\frac{\pi}{6} n_i \}_{n_i=0}^{n_S-1}$ are the coordinates of the minima. The three-dimensional vector of the dipole moment $\vecp$ in the minima deflects from the $ab$ plane by an angle $\theta$
\begin{equation}
\vecp_i = p_0 \cos(\theta)\vS_{i} \,,
\end{equation}
where $p_0$ is the magnitude of the dipole moment of the water molecule.
Here we introduced a new dimensionless vector variable
\begin{equation}\label{sigmai}
\vS_i =\left(\cos\frac{2\pi n_i}{n_S},\sin\frac{2\pi n_i}{n_S},(-1)^{z+n_i} \tan \theta\right)\,,
\end{equation}
where $z$ denotes the coordinate along the crystallographic $c$ axis.
Index $i$ should be considered a 3-dimensional integer variable, $i=\{x,y,z\}$ along axes $a,b,c$, respectively.
The interacting part of the full Hamiltonian can now be represented in the form of the classical Heisenberg spin model with spin vector variables $\vS_i$
\begin{equation}
    H_{I}=\frac{J}{2}\sum_{i\neq j}\vS_i\cdot \vecD_{i j}\cdot\vS_j - p_{\perp} \sum_{i}\vecB E_i\cdot \vS_i \,,
\end{equation}
where $J\equiv\bar{J}\cos^2\theta=p_0^2 \cos^2\theta/(4\pi \epsilon_0 \epsilon_r {r_0}^3)$ and $ p_{\perp}=p_0
\cos \theta$. The deflection angle $\theta=18.1\degree$ is the same for all dipoles.

\begin{figure}
\begin{center}
\resizebox{0.9\columnwidth}{!}{\includegraphics{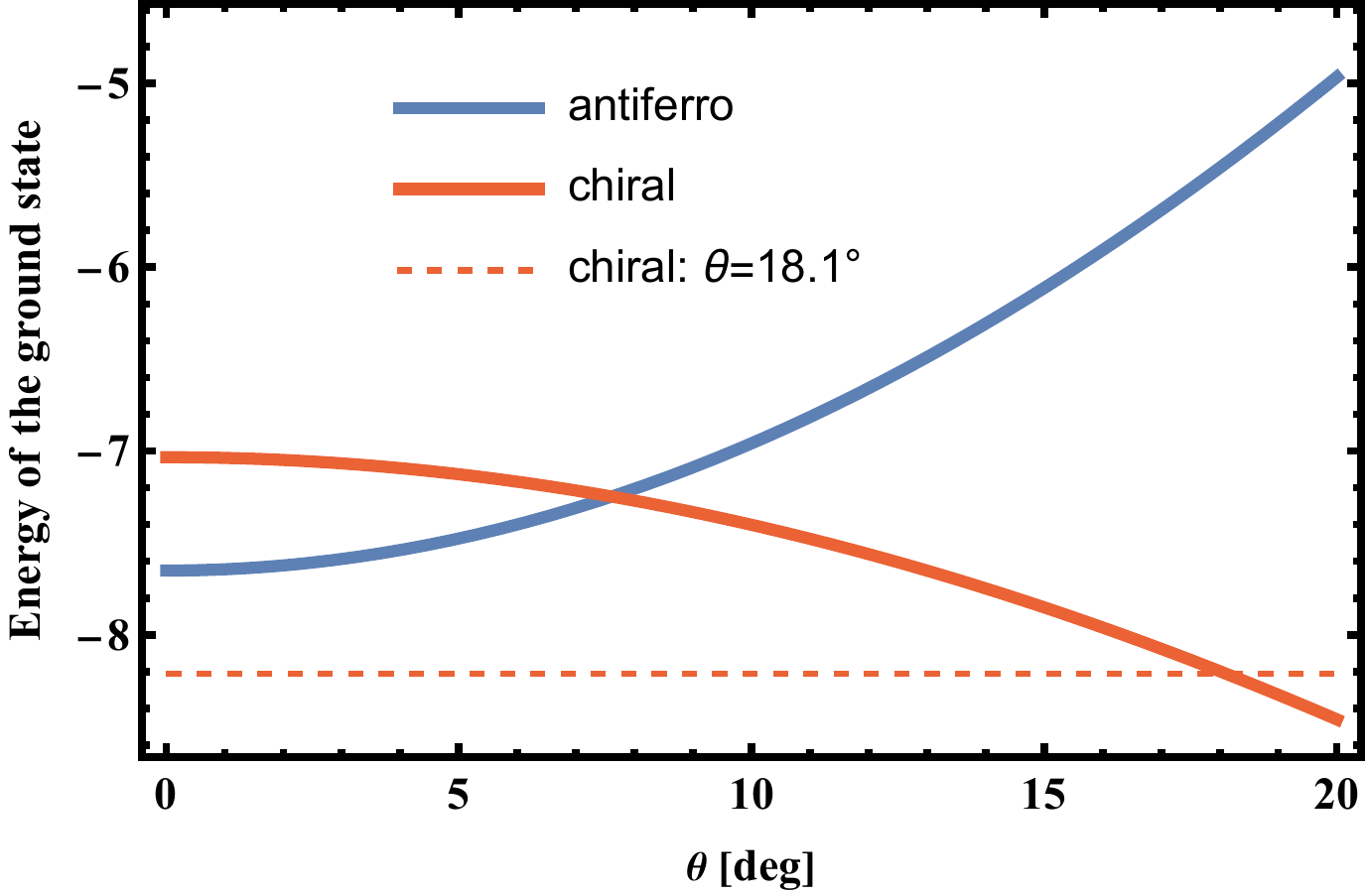}}
\end{center}
\caption{Comparison of energies of the two expected ground-state structures. Antiferroelectrically ordered ferroelectric planes vs. a chiral structure with twelve positions along the crystallographic $c$ axis as a function of the dipole deflection angle $\theta$ from the $ab$ plane are plotted. Energy is calculated in units where $\bar{J}=1$.}
\label{figGroundstateEnergy}
\end{figure}

\begin{figure}
\begin{center}
\resizebox{0.6\columnwidth}{!}{\includegraphics{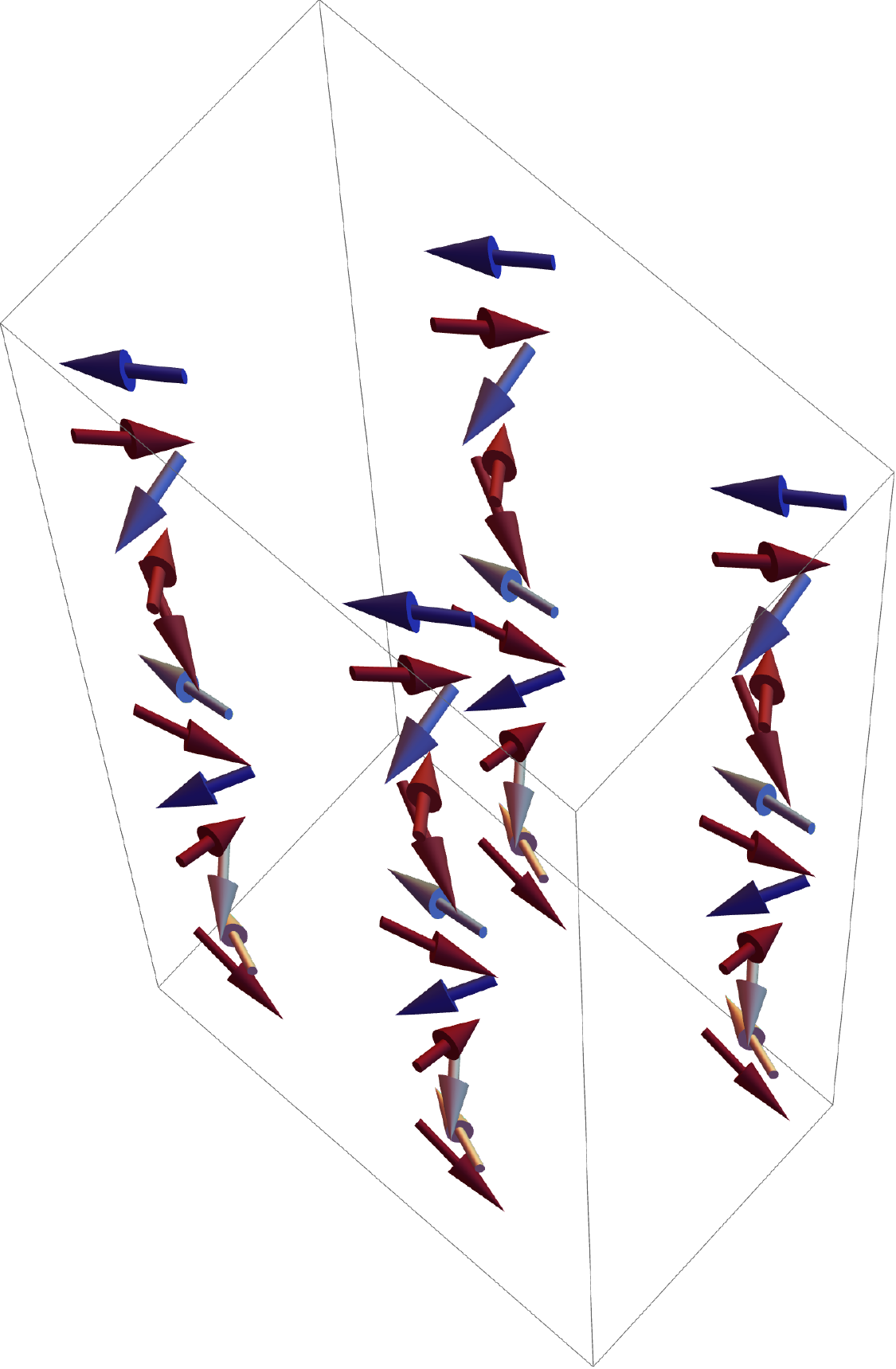}}
\resizebox{0.1\columnwidth}{!}{ }
\resizebox{0.18\columnwidth}{!}{\includegraphics{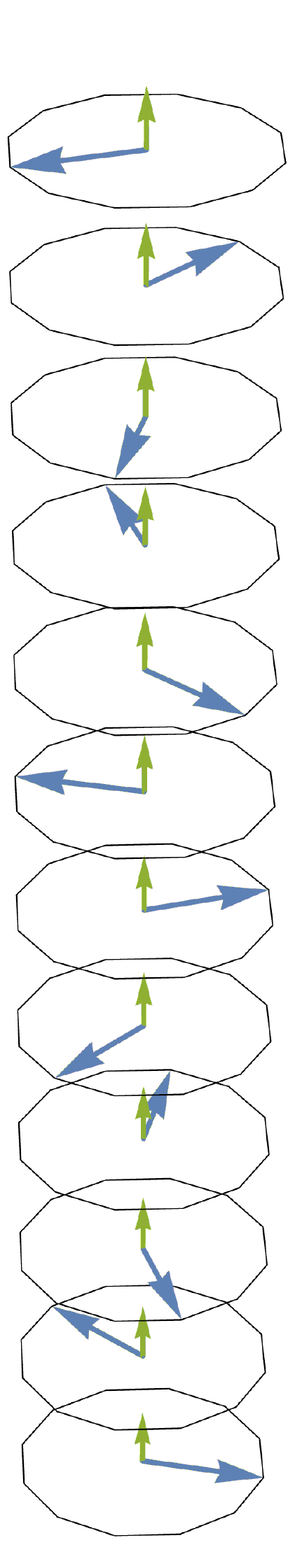}}
\end{center}
\caption{Visualized chiral structure of the dipoles ordered along the $c$ axis in the
ground state. On varying the $z$-coordinate, the dipoles go through all the twelve
minima of the local potential either clockwise or anti-clockwise. Left: four unit cells with the planar ($ab$) hexagonal structure. Right: a single unit cell with the dipole projections onto the $ab$-plane (blue arrows) and $c$-axis (green arrows).}
\label{figGroundstate}
\end{figure}

Finding the zero-temperature energy of any periodic configuration of the normalized polarization vectors $\vS_{i}$ is straightforward. Angle $\theta$ is a free parameter. The ground-state energy is obtained from minimizing a binary form
\begin{equation}
\mathcal{E}\left(\{\vS_{i}\}\right) = J(\theta)\sum_{j}\vS_{0}\cdot\vecD_{0j}\cdot\vS_{j}
\end{equation}
with respect to dipole configurations, depending on $\theta$. An ordered configuration gives the lowest
energy. It is, however, anisotropic and depends on the deflection angle
$\theta$. The dipole-dipole interaction drives the system to the saturated
planar ferroelectric order within the $ab$ planes. The order along the
$c$ axis is antiferroelectric for small
deflection angles $\theta<8\degree$, see Fig.~\ref{figGroundstateEnergy}.
The ground state displays a chiral structure for larger angles.
In the chiral structure, the dipoles are ferroelectrically ordered
within each $ab$ plane, but their angular coordinate $\phi$ switches to the
nearest minimum in one direction (either clockwise or anti-clockwise)
along the $z$ coordinate till it goes through the whole circle. It means that twelve coupled states are periodically repeated along the symmetry axis.  The unit cell of this helical structure is graphically presented in Fig.~\ref{figGroundstate}.

\subsection{Variational mean-field approximation}

The ground-state analysis shows the correct way of selecting the
unit cell of the periodic structure used in the thermodynamic description
of the water molecules in the beryl crystal. We start with a mean-field
description neglecting low-temperature quantum tunneling between the
minima of the degenerate ground state. The unit cell will have an internal
structure with twelve water molecules in the vertical direction.   The vector
order parameter, the thermally averaged value of the dipole moment
$\vecp(z)\sim\langle \vS_{z}\rangle$ contains index  $z=0,\ldots 11$ labeling the vertical coordinate within the unit cell conformal with the helical symmetry of the ground state.

We use the variational mean field and introduce a variational vector parameter $\vL_{z}$ with periodicity $n_S$ ($\vL_{z}=\vL_{z+n_S}$) that enters into the unperturbed Hamiltonian via
\begin{equation}\label{eqH0}
 H_0(\vL) = -\sum_{\iota,z} \vL_{z}\cdot\vS_{\iota,z} \,,
\end{equation}
where $\iota = \{x,y\}$ is the lattice index in the $ab$ plane.

We add the interacting part and treat it as a correction to the unperturbed one,
\begin{multline}
H_1\equiv H_{I}- H_0
\\
=\frac{J}{2}\sum_{i,j}\vS_i\cdot\vecD_{i j}\cdot\vS_j-\sum_{\iota}(\vecB{E} - \vL_{z})\cdot\vS_{\iota,z} \,.
\end{multline}
The variational free energy
\begin{equation}
F_{var}(\vL)\equiv F_0(\vL) + \langle H_1(\vL)\rangle_0
\end{equation}
is minimized to determine the variational parameters $\vL_z$.
 The unperturbed free energy is denoted $F_0$, and the unperturbed average is
\begin{equation}
\left\langle X\right\rangle_{0}\equiv\frac1{Z_{0}} \texttt{Tr}Xe^{-\beta H_{0}}\,,
\end{equation}
with the unperturbed partition sum
\begin{equation}
Z_0\equiv \texttt{Tr}e^{-\beta H_0}=\left( \prod_{z=0}^{n_S-1}\zeta_{z}(\vL_z)\right)^{N/n_S}\,.
\end{equation}
The vector with the independent self-consistently determined  parameters is $\vL_{z}= \{\lambda_{z}^{x},\lambda_{z}^{y},\lambda^z\}$.
The partition sum of the unit cell with the fixed vertical coordinate $z$ reads
\begin{equation}\label{eqZetaCl}
\zeta_{z}(\vL_z)= \sum_{l=0}^{n_S-1} e^{\beta \vL_z \cdot \vS_{z}(l)}
 \,.
\end{equation}
where $\vS_{z}(l)$ is defined in the same way as in Eq.~\eqref{sigmai}, $\vS_{z}(l) \equiv \{\cos(2\pi l/n_S),\sin(2\pi l/n_S), (-1)^{l + z}\tan\theta \}$.

The thermal average of the non-local interaction term in $H_1$ in the canonical ensemble with respect to $H_0$ is standardly decoupled, $\langle \vS_i\vS_j \rangle_0 =\langle \vS_i\rangle_0\langle\vS_j \rangle_0$.
Moreover, $\langle \vS_i\rangle_0$ depends not on the index $\iota$ ($x,y$ components of index $i$) but only on the $z$ component. We denote it by $\vP_z(\vL_z)$ and it can be expressed as
%
\begin{multline}\label{eq:Piz}
\langle \vS_z\rangle_0(\vL_z)=\frac{1}{\beta}\frac{\partial}{\partial \vL_{z}}\log \zeta_{z}(\vL_{z}) \\
=\frac{1}{\zeta_{z}(\vL_z)}
\sum_{l=0}^{n_S-1} \vS_{z}(l)e^{\beta \vL_z \cdot \vS_{z}(l)}\equiv \vP_z(\vL_z).
\end{multline}
The variational free energy density of the unit cell with $n_S$ water molecules along the $c$ axis $f_{var}\equiv F_{var}n_S/N$ can be written as
\begin{multline}
 f_{var}  = -\frac{1}{\beta}\sum_{z=0}^{n_S-1} \log \zeta_{z}(\vL_z)
\\
+\  \frac{J}{2}\sum_{z,z^{\prime}=0}^{n_S-1}\vP_z\cdot \vecD_{z z^{\prime}}\cdot \vP_{z^{\prime}}
- \sum_{z=0}^{n_S-1}(\vecB{E}_{z}- \vL_{z})\cdot\vP_z
\end{multline}
where
\begin{multline}
\vecD_{z z^{\prime}}\equiv\frac{N_z}{N}\sum_{\iota,\iota^{\prime}}\vecD_{\iota z, \iota^{\prime}z^{\prime}}
\\
=\frac{1}{N_z}\sum_{k_z}\tilde{\vecD}(k_x=0,k_y=0,k_z)e^{i k_z(z-z^{\prime})}\,.
\end{multline}
The $z$-component of the wave vector acquires integer multiples of $2\pi/n_S$ in
the dimensionless representation. We denote by $\tilde{\vecD}(\vecB{k})$ the dipole interaction in the momentum (wave vector) space
\begin{eqnarray}
\tilde{\vecD}(\vecB{k})\equiv\frac{1}{N}\sum_{i,j}\vecD_{ij}e^{-i \vecB{k}\cdot(\vecB{r}_i-\vecB{r}_j)} \,,
\end{eqnarray}
that is evaluated for a given structure of an infinite or periodic system by using the Ewald method, see, e.g., \cite{Rapaport2004}.

Stationary equations $\partial f_{var}/\partial \vL_{z}=0$ determine the equilibrium values of  the variational parameters $\bar{\vL}_{z}$, mean fields,
\begin{equation}
\bar{ \vL}_{z}=\vecB{E}_{z} - J\sum_{z^{\prime}=0}^{n_S-1}\vecD_{z z^{\prime}}\cdot\vP_{z^{\prime}}(\bar{\vL}_{z^{\prime}})\,.
\end{equation}

The corresponding conjugate equations for the order parameters, thermally
averaged local polarizations, read
\begin{equation}\label{eqMF}
\langle \vS_z\rangle_0=\vP_z\left(\vecB{E}_{z} - J\sum_{z^{\prime}=0}^{n_S-1}\vecD_{z z^{\prime}}\cdot\langle \vS_{z^{\prime}}\rangle_0\right) \,.
\end{equation}

\subsection{Spatial fluctuations: Monte Carlo simulations}

The mean-field approximation averages spatial fluctuations and assumes their
effect as a mean homogeneous environment, a mean field. The dipole-dipole
interaction has a long-range and anisotropic character, and replacing it with an
effective constant is a crude approximation. Such an approximation
generically overestimates the tendency toward a long-range order. Hopping and tunneling in the clock model only partly suppress this tendency. The spatial fluctuations significantly affect ordering, particularly in low spatial dimensions. We used the local clock model, and instead of using the mean-field solution for the crystal equilibrium state, we used the classical Metropolis Monte Carlo simulations \cite{Metropolis:1953aa,Frenkel:1990aa} and at low temperatures we also used the Wang-Landau (WD) flat histogram method \cite{WangLandau1,WangLandau2}.

\subsection{Low-temperature quantum tunneling}
\label{sec:QT}

The classical mean-field solution reflects well the high-temperature behavior of statistical models. Low-temperature asymptotics may, however, be
strongly affected by quantum fluctuations. In extreme situations, they can
destroy the classical behavior by the existence of quantum
criticality with quantum critical points. Very recently, a quantum critical
point was predicted in a linear chain of the dipole moments of the water
molecules \cite{Serwatka:2023aa}. We assess the impact of quantum fluctuations
within the clock model in a simplified form by allowing the hopping of the dipole moments only between the
nearest and next-to-nearest potential minima. The
hopping amplitudes $a$ and $d$ describe the
transitions between the minima from the opposite and the same oxygen
planes with and without changing the vertical
component of the dipole moment, respectively, as shown in
Fig.~\ref{fig:2x6-Potential} by the oblique and horizontal dotted arrows. The new
variational unperturbed Hamiltonian of the mean-field approximation with
quantum tunneling remains local, but it becomes non-diagonal
\begin{equation}\label{eqH0Q}
 \hat{H}_0(\vL) = \hat{h}^{loc} -\sum_{\iota,z} \vL_{z}\cdot\hat{\vS}_{\iota,z} \,,
\end{equation}
where $\hat{h}^{loc}$ is a $12\times 12$ matrix defined in Eq.~\eqref{eq:TwelveMatrix}.  The local mean-field partition function Eq.~\eqref{eqZetaCl} is then
\begin{equation}
\zeta(\vL_z)= \texttt{Tr} e^{-\beta \hat{h}^{loc}_{mf}}=\sum_{n=0}^{n_S} e^{-\beta v_n(z)}\,,
\end{equation}
where the $v_n(z)$ are eigenvalues of the matrix
\begin{equation}
\hat{h}^{loc}_{mf}= \hat{h}^{loc} - \left\{\delta_{ij} \vL_{z}\cdot\vS_{z}(j)\right\}_{i,j=1}^{n_S}  \,.
\end{equation}

The mean-field polarization is then defined from the local partition function in  Eq.~\eqref{eq:Piz}. The non-local interacting part is added in the same way as in the classical case to reach the corresponding mean-field solution with quantum tunneling.

\section{Thermodynamic properties}

\subsection{Mean field approximation}

\begin{figure}
\begin{center}
\resizebox{0.99\columnwidth}{!}{\includegraphics{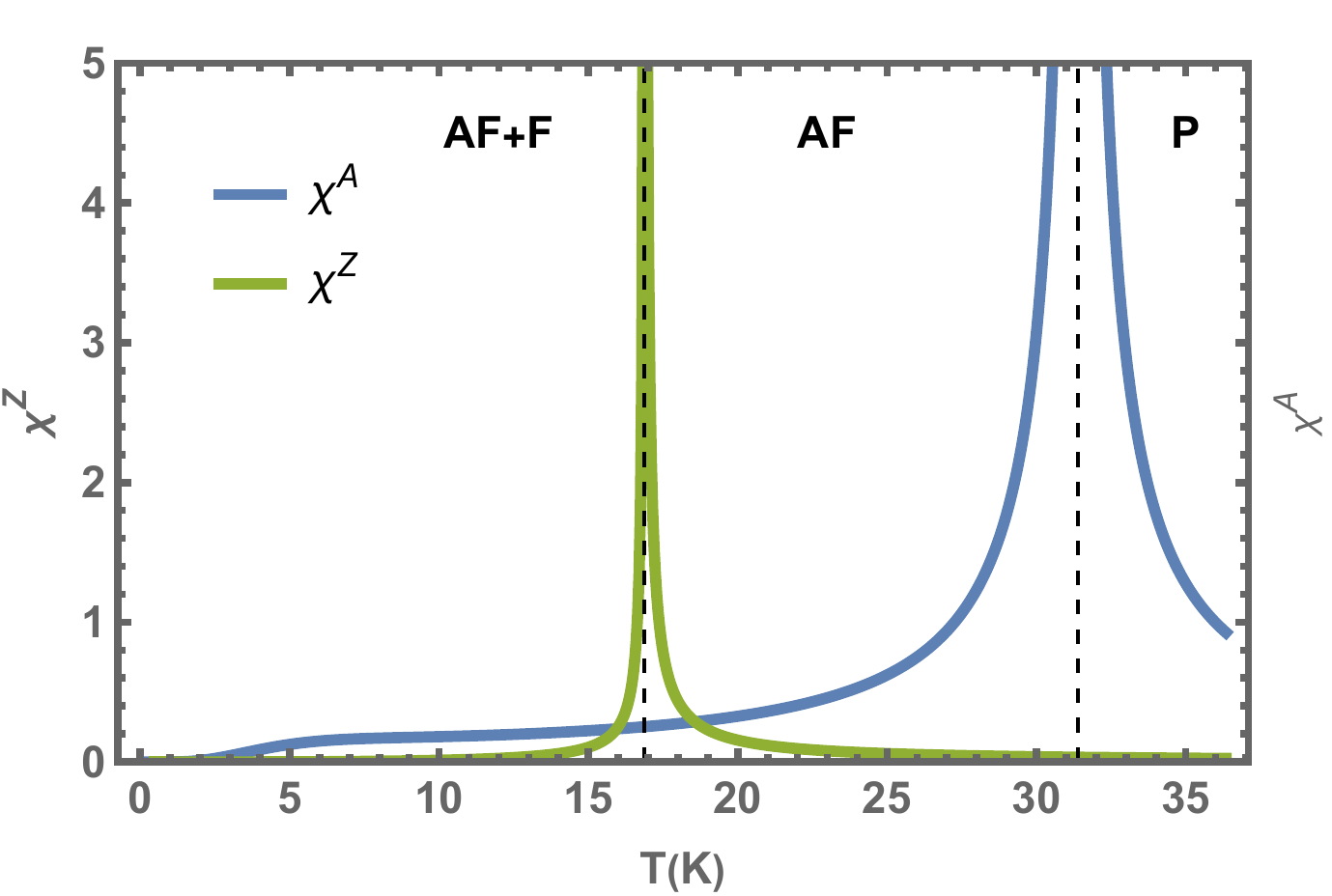}}
\end{center}
\caption{Staggered (along the $c$ axis) static susceptibility $\chi_A$ (electric field in $a$ direction), and static susceptibility $\chi_Z$ (electric field in $c$ direction) indicating the two critical points below which the planar, $T_{c1}$,  and vertical, $T_{c2}$,  orders emerge, respectively.}
\label{figSuscAsym}
\end{figure}

We learned from the ground-state calculations that we can expect two ordered
phases in the low-temperature limit.  Approaching this limit from high
temperatures, we start with the variational mean-field approximation with the
unit cell containing $n_{S}=12$ water molecules along the $c$ axis to
assess the trends in the thermodynamic behavior when decreasing temperature. The
first ordering emerges at $T_{c1}\sim 31.6\,\rm K$ below which the thermally
averaged moments are planar and ferroelectrically ordered in the individual
$ab$ planes. Due to the anisotropic character of the dipole-dipole interaction,
the ferroelectric planes order antiparallelly in the adjacent planes. The
response to the electric field along the $z$ axis increases with decreasing temperature and diverges at another critical point
$T_{c2}\sim 16.8\,\rm K$, Fig.~\ref{figSuscAsym}. Below $T_{c2}$, the
thermal average of the $c$ projection of the dipole moment becomes non-zero and
parallelly ordered along the $c$ axis.

The local minima of the single dipole moment are spread between two adjacent $ab$ planes. The dipole moments can be perfectly antiparallel in the adjacent planes only if they stay planar, $T_{c1}> T >T_{c2}$. Once the  $c$ projection of the dipole moment becomes macroscopically non-zero, the perfect antiparallel order is no longer possible, Fig.~\ref{figAntiFerroStates}. The planar component of the dipole moment starts to drift in one direction, either clockwise or anticlockwise.  It means that the $a$- and $b$-projections of the dipole moment are not the same when the $c$ projection is nonzero, Fig.~\ref{figSolutions}.
\begin{figure}
\begin{center}
\resizebox{0.2\textwidth}{!}{\includegraphics{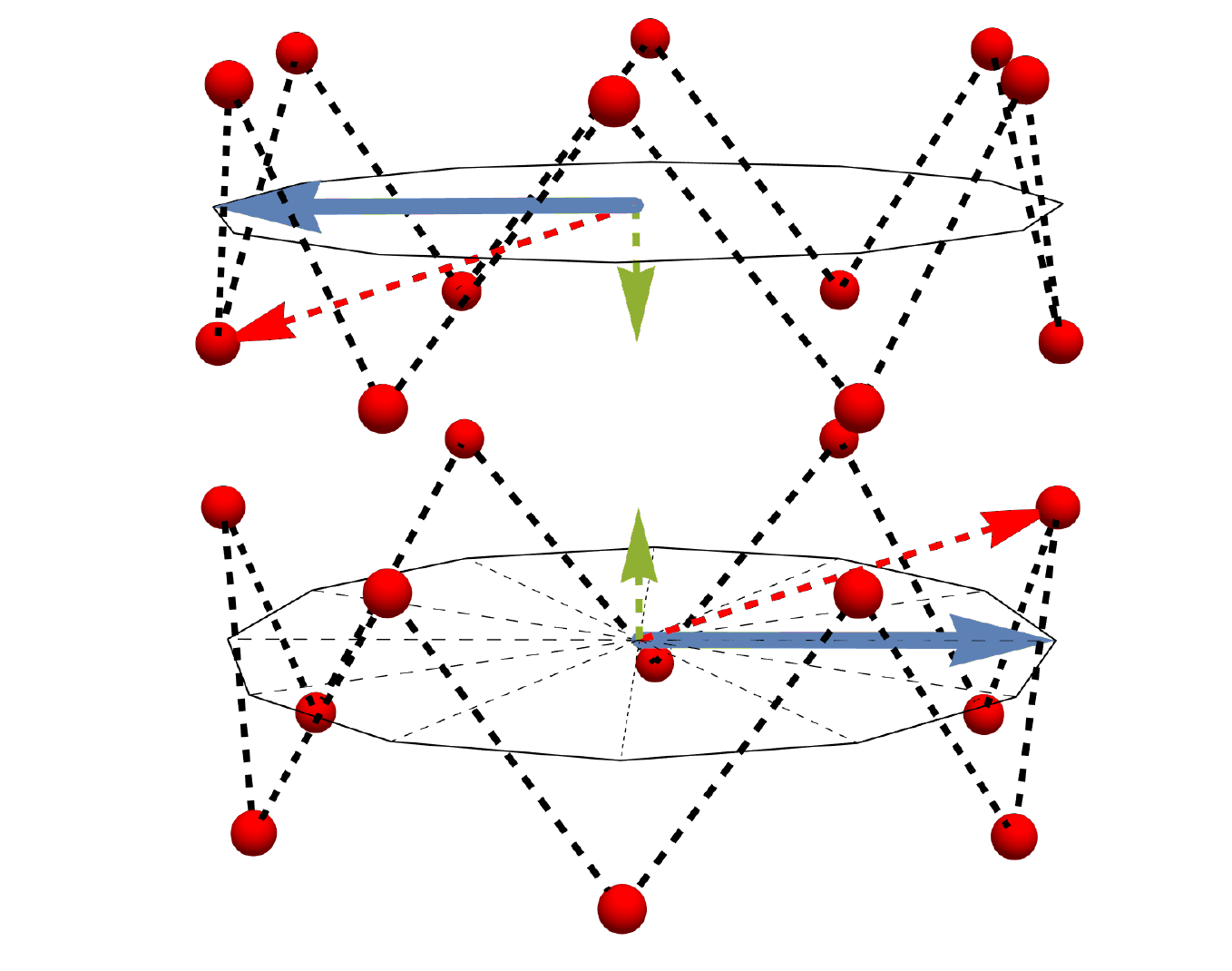}}
  \setlength{\unitlength}{2144sp}%
  \begingroup\makeatletter\ifx\SetFigFont\undefined%
  \gdef\SetFigFont#1#2#3#4#5{%
    \reset@font\fontsize{#1}{#2pt}%
    \fontfamily{#3}\fontseries{#4}\fontshape{#5}%
    \selectfont}%
  \fi\endgroup%
  \begin{picture}(489,527)(874,-1438)
  \thinlines
  {\color[rgb]{0,0,0}\put(901,-1411){\vector( 0, 1){450}}
  }%
  {\color[rgb]{0,0,0}\put(901,-1411){\vector( 2, 1){360}}
  }%
  {\color[rgb]{0,0,0}\put(901,-1411){\vector( 1, 0){450}}
  }%
  \put(1286,-1379){\makebox(0,0)[lb]{\smash{{\SetFigFont{10}{14.4}{\rmdefault}{\mddefault}{\itdefault}{\color[rgb]{0,0,0}a}%
  }}}}
  \put(1101,-1203){\makebox(0,0)[lb]{\smash{{\SetFigFont{10}{14.4}{\rmdefault}{\mddefault}{\itdefault}{\color[rgb]{0,0,0}b}%
  }}}}
  \put(929,-1016){\makebox(0,0)[lb]{\smash{{\SetFigFont{10}{14.4}{\rmdefault}{\mddefault}{\itdefault}{\color[rgb]{0,0,0}c}%
  }}}}
  \end{picture}
\resizebox{0.2\textwidth}{!}{\includegraphics{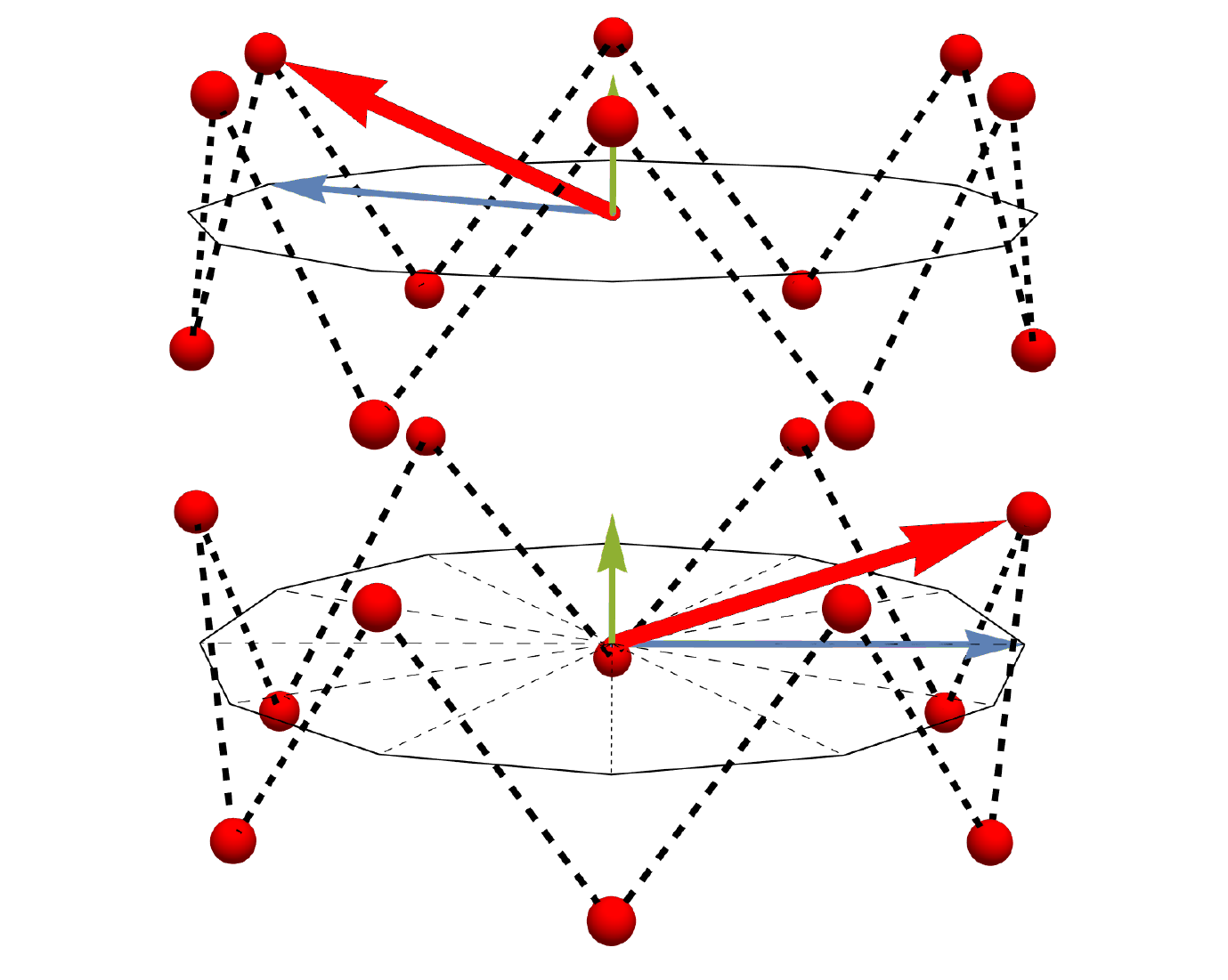}}
\end{center}
\caption{Left pane: Ferroelectric order emerges in the $ab$ planes
below $T_{c1}$, with the planes alternating the direction of the dipole moment, forming an antiferroelectric structure along the $c$ axis. The instantaneous directions of the dipole
moments (red arrows) deflect from the $ab$ planes, but the thermal
average of the $c$ components (green arrows) is zero for $T>T_{c2}$.
Right pane: Configuration of the water molecules for $T<T_{c2}$ with long-range
ferroelectric order of the $c$ projections of the dipole moments. The
vertical components of the dipole moments are aligned
parallel, whereas the $ab$-plane components
(blue arrows) in the neighboring planes are one angular step away
from the prohibited antiparallel orientations.}
\label{figAntiFerroStates}
\end{figure}
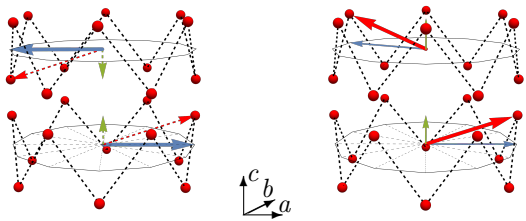
\begin{figure}
\begin{center}
\resizebox{0.99\columnwidth}{!}{\includegraphics{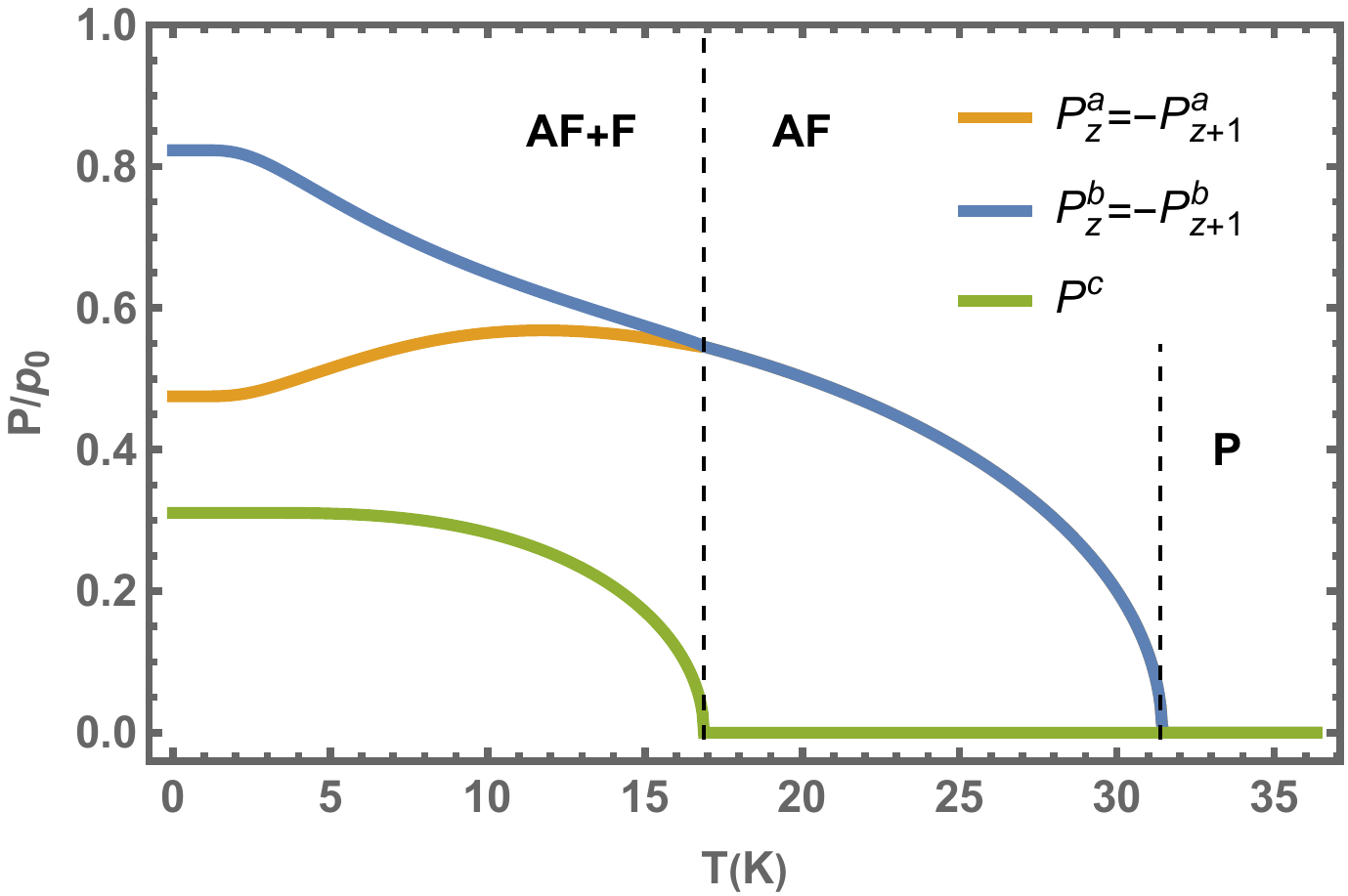}}
\end{center}
\caption{Three thermodynamic phases are observed in the mean-field solution of the
water molecules in the beryl cavities. At high temperatures ($T>T_{c1}$), the system is in the paraelectric (P) phase, which lacks long-range order.
At intermediate temperatures ($T_{c1}>T>T_{c2}$), the system exhibits planar ferroelectric order with antiparallel alignment along the $c$ axis (AF).
At low temperatures ($T<T_{c2}$), the system exhibits long-range order with all three projections nonzero and ferroelectric order (F) along the $c$ direction. In this phase, the splitting of the $a$- and $b$-projections is clearly evident.}
\label{figSolutions}
\end{figure}

The projection of the dipole moment in the $ab$-plane changes its direction when moving along the $c$ axis.
We take the unit cell with twelve water molecules in the $c$ direction to characterize these changes. We define an order parameter for this unit cell, denoted by $\vecP$, which reflects the antiferroelectric order in the $ab$-plane along the $c$-axis and the ferroelectric order in the $c$-axis. The parameter is given by:
\begin{equation}
\vecP \equiv\frac{1}{n_S} \sum_{z=0}^{n_S-1} \left\{(-1)^z p^x(z),(-1)^z p^y(z), p^z(z)\right\} \,
\end{equation}
where $n_S=12$ represents the number of molecules in the unit cell. This parameter is referred to as the antiferroelectric/ferroelectric (AF/F) order parameter.

\begin{figure}
\begin{center}
\resizebox{0.99\columnwidth}{!}{\includegraphics{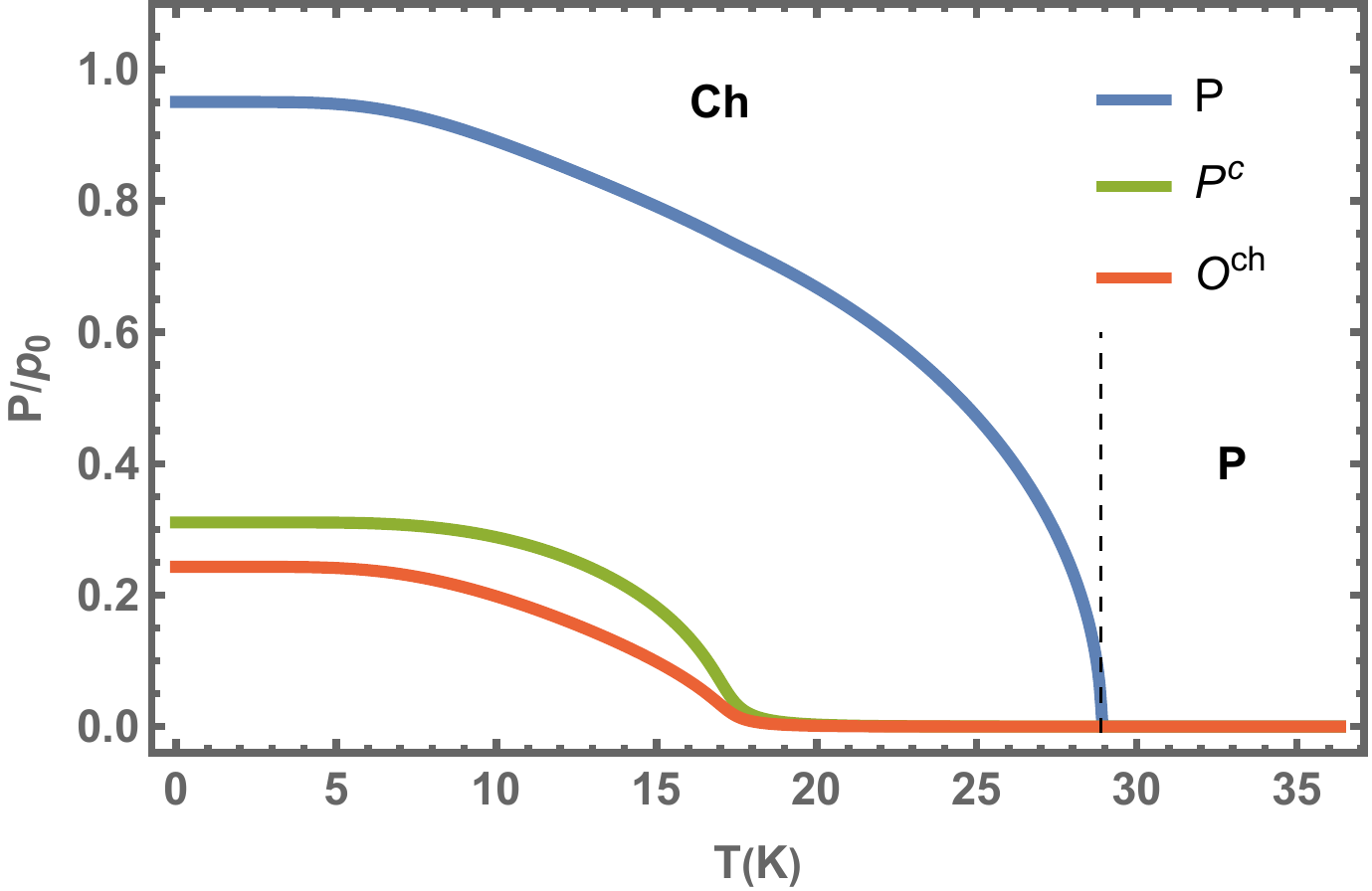}}
\end{center}
\caption{Three thermodynamic order parameters connected with antiferroelectrically ordered planes, $P$, $c$ projection of the dipole moment, $P^{c}$, and the chiral parameter, $O^{ch}$. }
\label{figSolutions-chiral}
\end{figure}

The modulus of this thermally averaged vector parameter is nonzero below $T_{c1}$, and the vector remains planar for $T>T_{c2}$. To understand what happens below $T_{c2}$, we need to introduce another vector parameter that can distinguish a drift of the planar projection of the dipole moment in one direction from random fluctuations when moving along the $c$ axis.  If one direction of the drift is preferred, we end up with a chiral order. To assess the chiral order quantitatively, we introduce a parameter $O^{ch}$ using polarizations in three consecutive layers \cite{Tsvelik2017}
\begin{equation}
\mathbf{O}^{ch} = \frac{1}{n_S} \sum_{z=0}^{n_S-1}  \vecp_z\cdot(\vecp_{z+1}\times \vecp_{z+2})\,.
\end{equation}
This order parameter becomes non-zero whenever $P^{c}\neq 0$ as demonstrated in
Fig.~\ref{figSolutions-chiral}. The chiral phase is a local minimum of
the free energy for $T<T_{c2}$.  The thermally averaged parameter $\mathbf{P}$ has nonzero only the $c$-component below $T_{c2}$ in the chiral state.
%
%

%

\begin{figure}
\begin{center}
\resizebox{0.99\columnwidth}{!}{\includegraphics{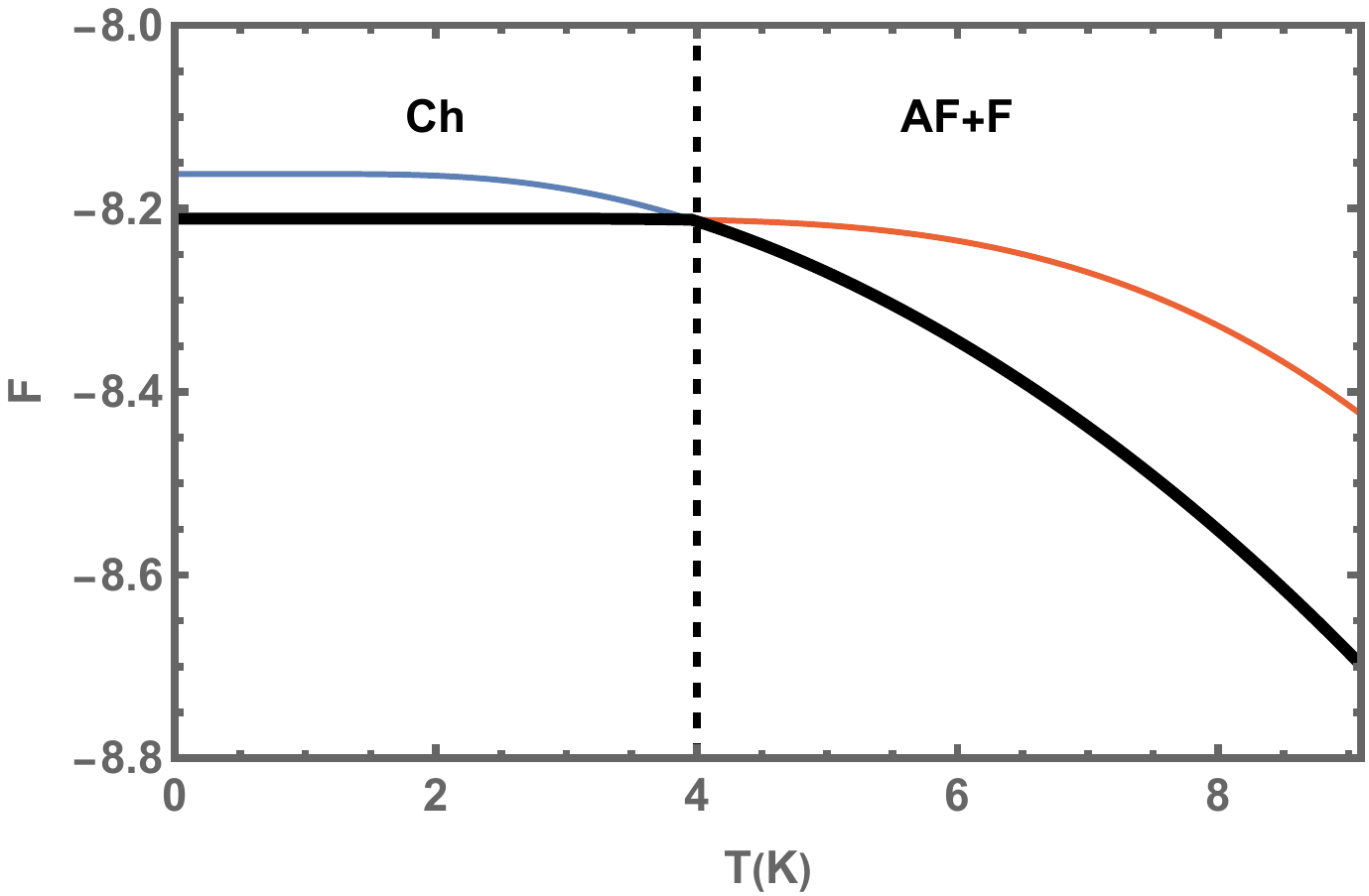}}
\end{center}
\caption{Temperature dependence of free energy of the antiferroelectric and chiral ordering of the dipole moments along the $c$ axis. They cross at a first-order phase transition point, $T_0\simeq4 K$. The black line is the equilibrium free energy.}
\label{figMFfreeEnergy}
\end{figure}

\begin{figure}
\begin{center}
\resizebox{0.99\columnwidth}{!}{\includegraphics{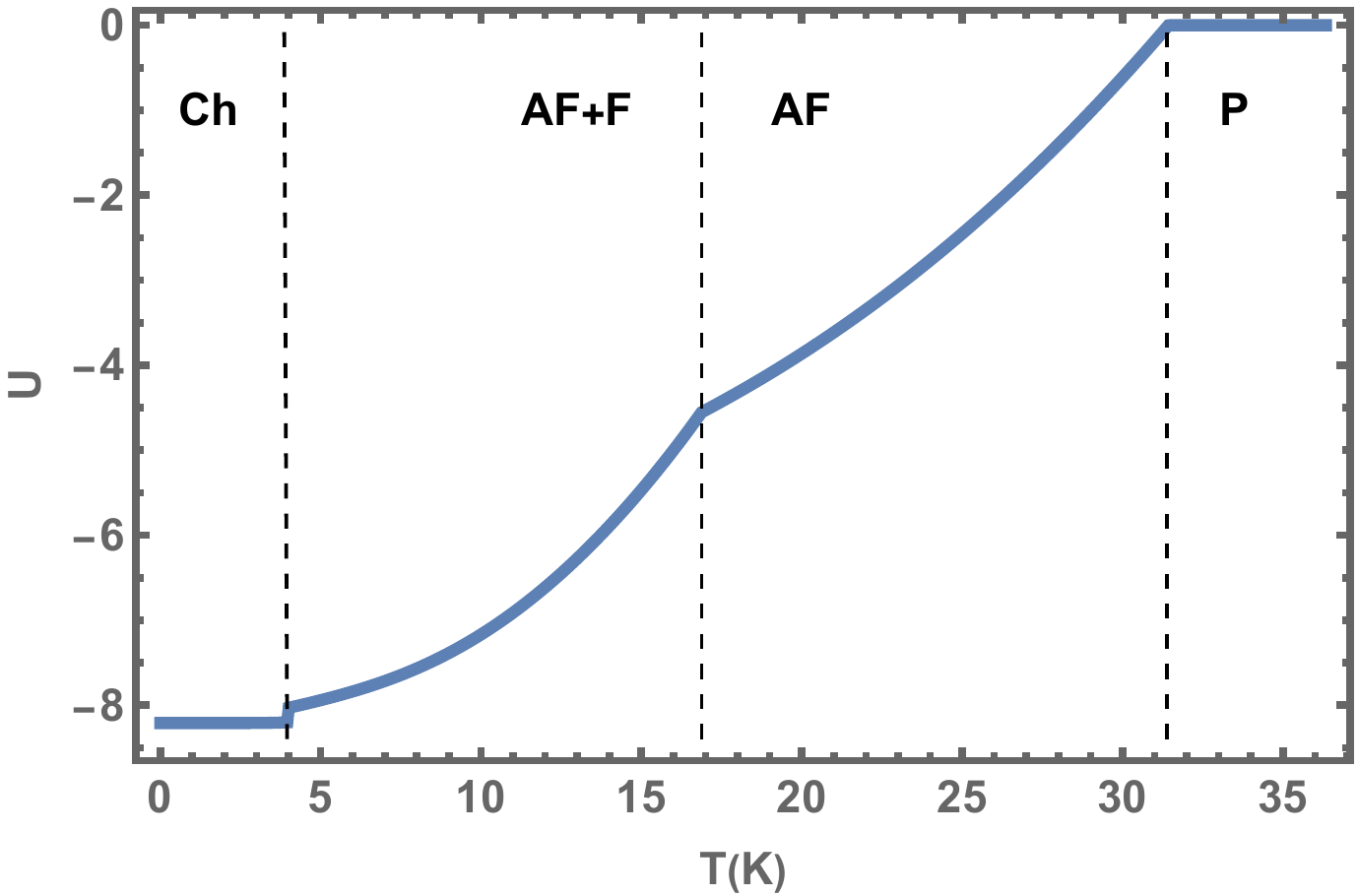}}
\resizebox{0.99\columnwidth}{!}{\includegraphics{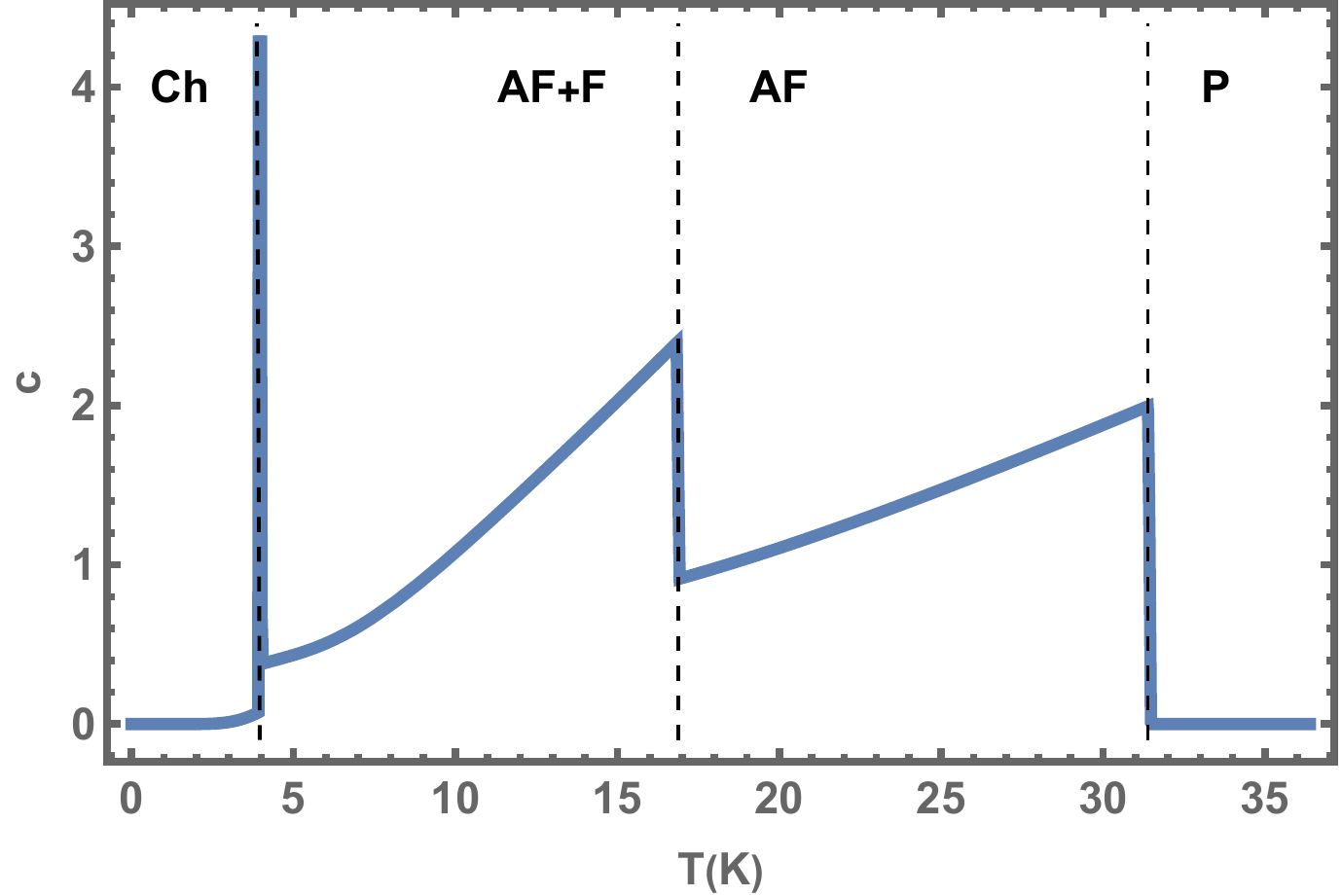}}
\end{center}
\caption{Temperature dependence of the mean-field internal energy (upper figure) and heat capacity (lower figure) indicating the three transition points. The upper two cusps of heat capacity correspond to continuous transitions at temperatures $T_{c1}$ and $T_{c2}$, and the lowest to a discontinuous transition at $T_0$.}
\label{figCapacityMF}
\end{figure}

\begin{figure}
\begin{center}
\resizebox{0.99\columnwidth}{!}{\includegraphics{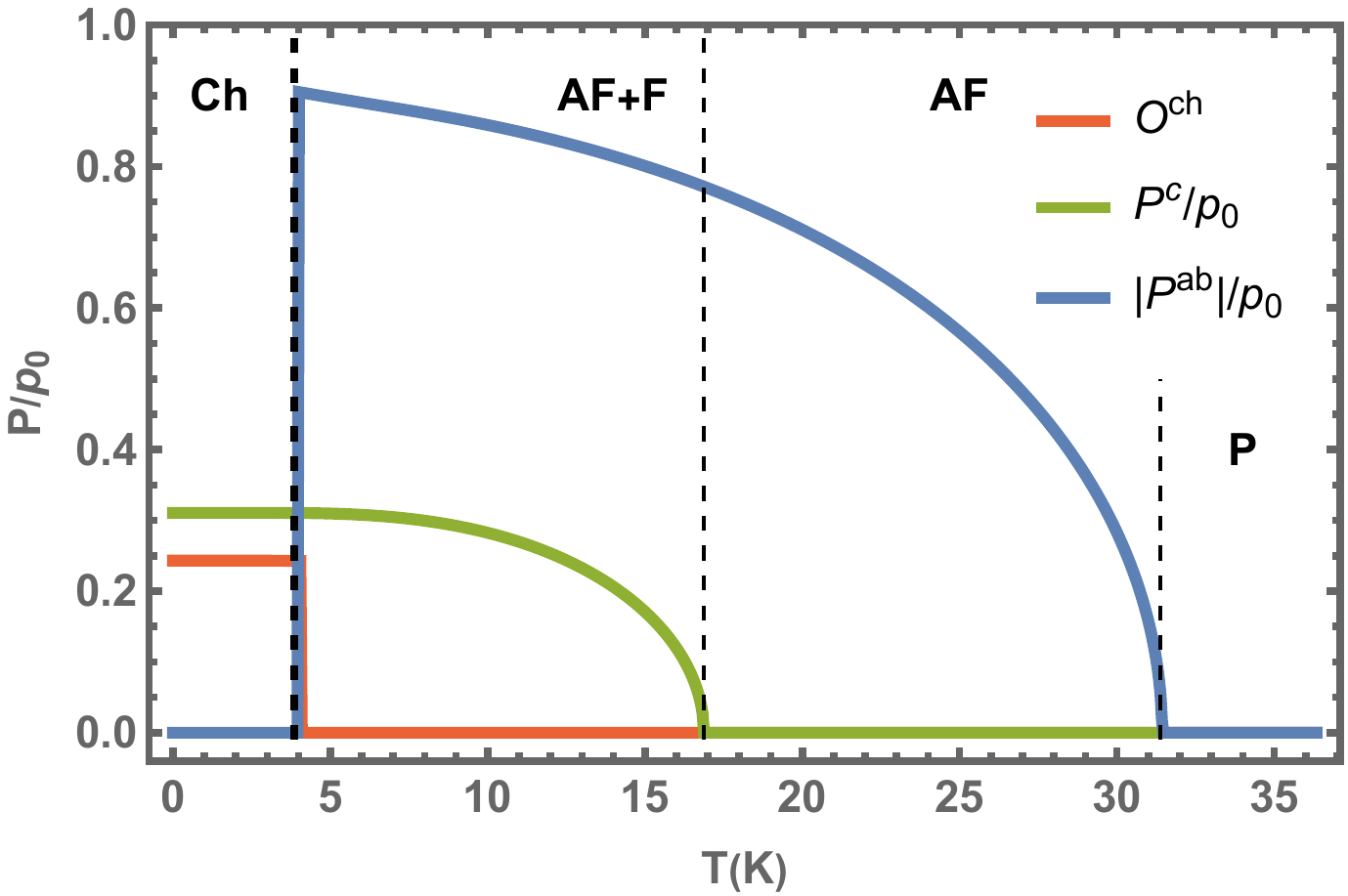}}
\end{center}
\caption{Temperature dependence of the mean-field order parameters as discussed in the text calculated in the true equilibrium state. Parameter $\vecP^{ab}_{A/F} = 0$ in the chiral state (Ch) when $O^{Ch}$ saturates.}
\label{figMFPolarizationC}
\end{figure}

We see that below the critical point $T_{c2}$, at which the vertical component of the thermally averaged dipole moment becomes non-zero, both the unit-cell antiferroelectric order parameter
$\mathbf{P}$ and chiral order parameter $\mathbf{O}^{ch}$ can be
non-zero in equilibrium. These parameters describe different thermodynamic states that generally differ in free energy. However, only one of the solutions represents the true equilibrium state. The free energies in the mean-field approximation are plotted
in Fig.~\ref{figMFfreeEnergy}. The chiral state has a lower free energy below
$T_{0}\sim 4\,\rm K$, at which a first-order transition to equilibrium takes place.
This conclusion is supported by the internal energy and heat capacity, Fig.~\ref{figCapacityMF}.
The corresponding behavior of the thermodynamic order parameters is plotted in Fig.~\ref{figMFPolarizationC}.
%

%

\subsection{Monte Carlo simulations}

The variational mean-field approximation suppresses spatial fluctuations and
replaces the crystal with a mean environment. To improve this simplification,
we used Monte Carlo simulations to take the spatial fluctuations into
account more sophisticatedly.

\begin{figure}
\begin{center}
\resizebox{0.99\columnwidth}{!}{\includegraphics{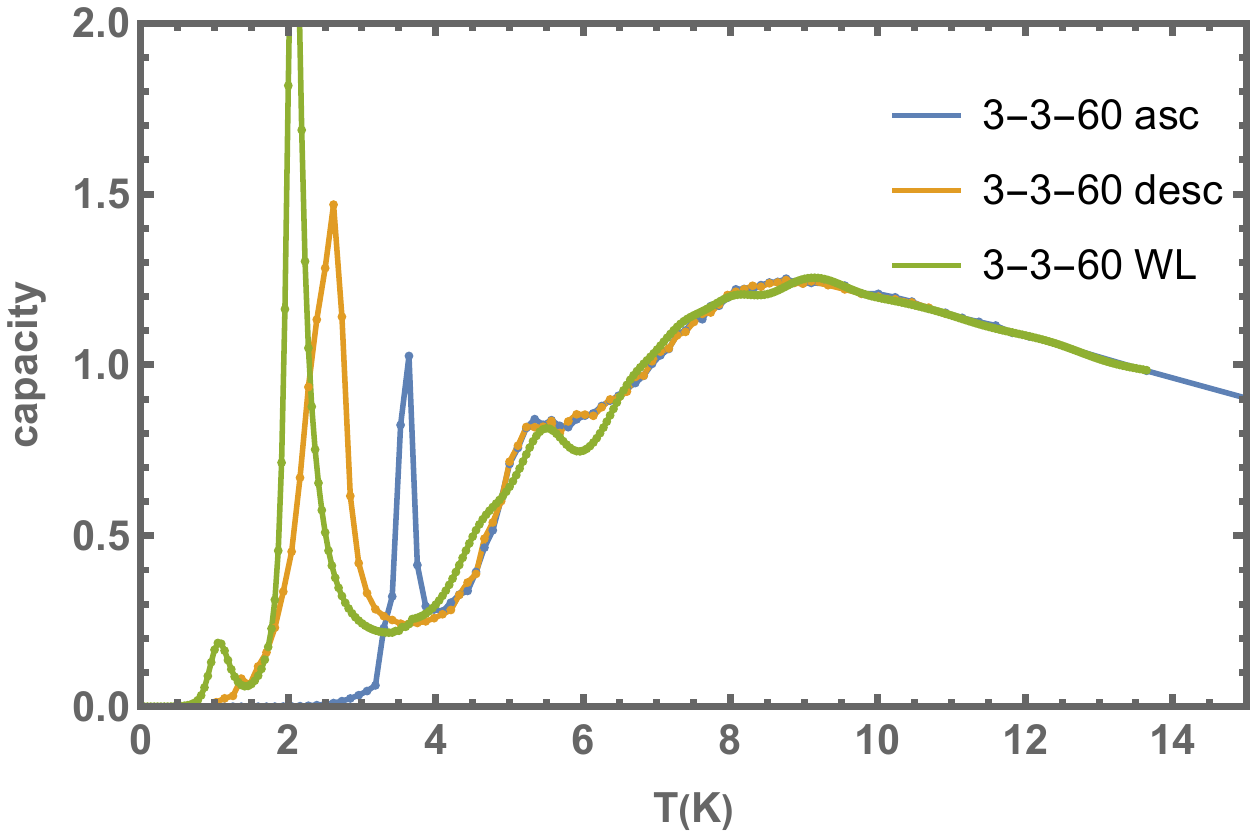}}
\end{center}
\caption{Temperature dependence of the Monte Carlo heat capacity of the water dipole moments with three transitions to ordered phases. The heat capacity exhibits a cusp at around $5.5\,\rm K$. Different ways for calculating the heat capacity were employed, including descending (desc) and ascending (asc) temperatures as well as the Wang-Landau method. The first two direct approaches show hysteresis with two transition temperatures. In contrast to the mean-field solution,  the Monte Carlo simulations suggest a first-order transition to the phase with ferroelectrically ordered $c$-projections of the polarization.}
\label{figCapacityMC}
\end{figure}

\begin{figure}
\begin{center}
\resizebox{0.99\columnwidth}{!}{\includegraphics{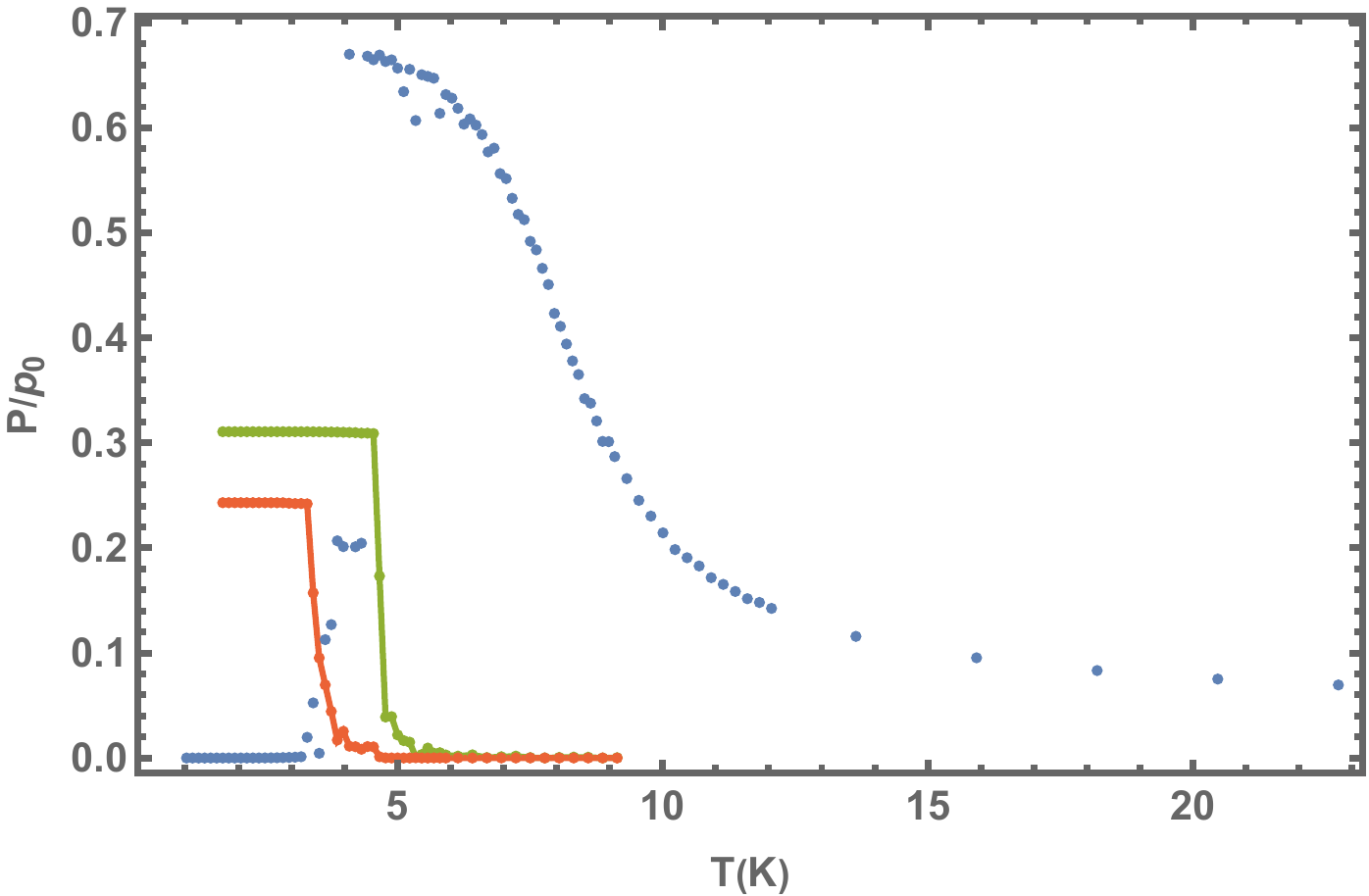}}
\end{center}
\caption{
Temperature dependence of the Monte Carlo polarizations when the system is heated from the ground state. Planar order antiferroelectrically aligned along the $c$ axis (blue dots), ferroelectric order of the $c$ projection of the dipole moment (green), and the helical order parameter (red). We can observe that the effective dipole moment's planar projection in the unit cell with $n_{S}=12$ water molecules becomes zero in the helical state.}
\label{figPolarizationMC}
\end{figure}

We employed the standard Metropolis dynamics with single-site updates and improved them with non-local updates at low temperatures to explore our system's phase space more effectively.
We investigated the system with linear size in $a$ and $b$ directions of $L_a=L_b \in \{3,4,5,6\}$ and up to 8 cells of 12 atoms in the $c$ direction, assuming periodic boundary conditions.
Averages were calculated using at least $2\times10^6$ Monte Carlo steps (MCS) per dipole after discarding other $1\times10^6$ MCS for equilibration. At low temperatures, up to $5\times10^6$ MCS were used.
The heat capacity calculated with the Monte Carlo simulations is plotted in Fig.\ref{figCapacityMC}. The broad peak in the plot indicates the second-order phase transition to the planar order at $T_{c1} \approx 9.5\, \rm K$. We observe a cusp at  $ \approx 5.5\,\rm K$ indicating a non-analytic behavior connected with the emergence of ferroelectrically ordered chains of the dipoles thermodynamically averaged only along the $c$-axis. The vertical chains are, however, not ordered when changing their $ab$-coordinates.  The convergence significantly slows down when decreasing temperature. Two metastable states of the $c$-ordered chains coexist on long-time scales. One is the bulk ferroelectric state, and the other is a state with disordered ferroelectric chains. The whole system gets stuck in one of these metastable states during the Monte Carlo simulations. The time needed to overcome the barrier between these states grows exponentially at very low temperatures. We, therefore, alternatively used the Wang-Landau (WD) flat histogram method \cite{WangLandau1,WangLandau2} to handle this problem. It allowed us to calculate the density of states, from which we could directly deduce the temperature dependence of the free energy, internal energy, and heat capacity.  Based on this, we identified a transition at $T_{c2} \approx 2\,\rm K$ to a bulk ferroelectric order where the ferroelectric  $c$-chains are aligned parallelly.  A transition to the helical state was observed at $T_{0} \approx 1\,\rm K$. Sharp heat capacity peaks indicate a discontinuity in the internal energy at both transitions. This conclusion was confirmed by discontinuities in the polarization shown in Fig.~\ref{figPolarizationMC}. The Monte Carlo simulations indicate that the phase transitions to the bulk ferroelectric ordering of the $c$-chains at $T_{c2}$ and to the helical state at $T_{0}$ are of first-order type. The transition to the ferroelectric order along the $c$-axis is second order in the mean-field approximation.

The precise determination of the transition temperature $T_{c2}$ from Monte Carlo simulations becomes difficult due to the coexistence of two metastable states. We observed a hysteresis depending on whether we approached the transition from above or from below. When starting from the ordered low-temperature state, the transition temperature with a discontinuity in the internal energy is higher than that approached from the high-temperature disordered phase. This hysteresis is, however, suppressed by the Wang-Landau method, Fig.~\ref{figCapacityMC}. It is worth mentioning that the cusp near $5.5\,\rm K$ is invisible in systems with only a few cells along the $c$-axis. Long-range spatial fluctuations generally pull down the transition temperatures. The Monte Carlo simulations with spatial fluctuations support, nevertheless, the overall picture of the mean-field solution. They additionally predict the existence of inhomogeneous metastable states, possibly leading to a first-order transition to the state with a non-zero projection of the polarization to the $c$-axis. A more detailed analysis of the model with Monte Carlo simulations will be published separately.

\subsection{Quantum tunneling}

\begin{figure}
\begin{center}
\resizebox{0.99\columnwidth}{!}{\includegraphics{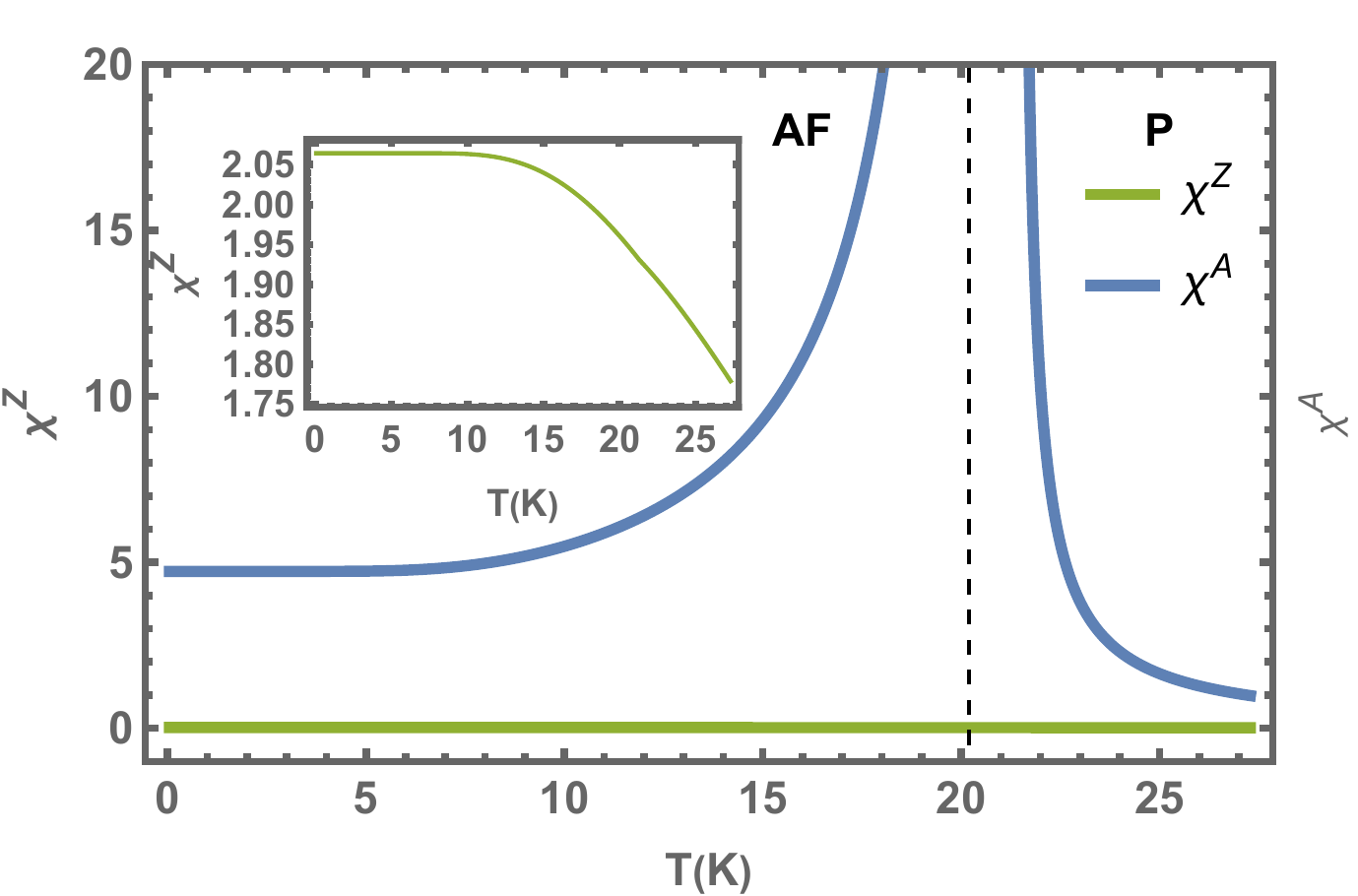}}
\resizebox{0.99\columnwidth}{!}{\includegraphics{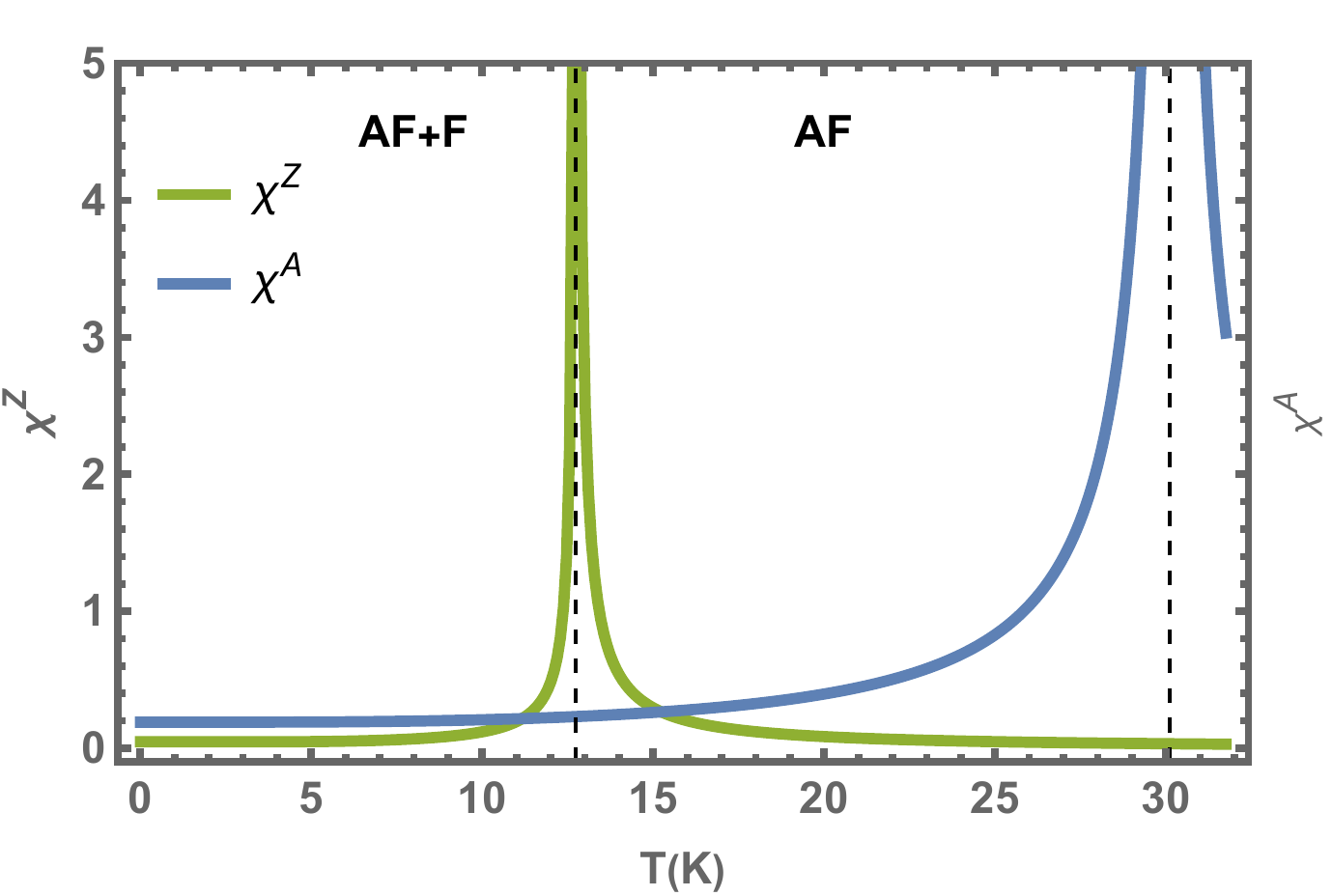}}
\end{center}
\caption{Antiferroelectric (blue) and ferroelectric (green)
susceptibilities calculated with quantum tunneling amplitudes as discussed
in Sec.~\ref{sec:QT}. Dynamical quantum fluctuations decrease the
transition temperatures, as do the static spatial fluctuations. The
quantum tunneling for setting labeled "1" (with $V_{b1},\omega_1$) appears to lead to a zero $c$-projection of the thermally averaged dipole moment (top figure). The inset shows the ferroelectric susceptibility with a magnified vertical scale.
The second setting (bottom figure) labeled "2" (with $V_{b2},\omega_2$) allows for the existence of vertical ferroelectric order at a non-zero temperature  $T_{c2}$, although with suppressed chiral order.}
\label{figMFPolarizationQ}
\end{figure}

\begin{figure}
\begin{center}
\resizebox{0.99\columnwidth}{!}{\includegraphics{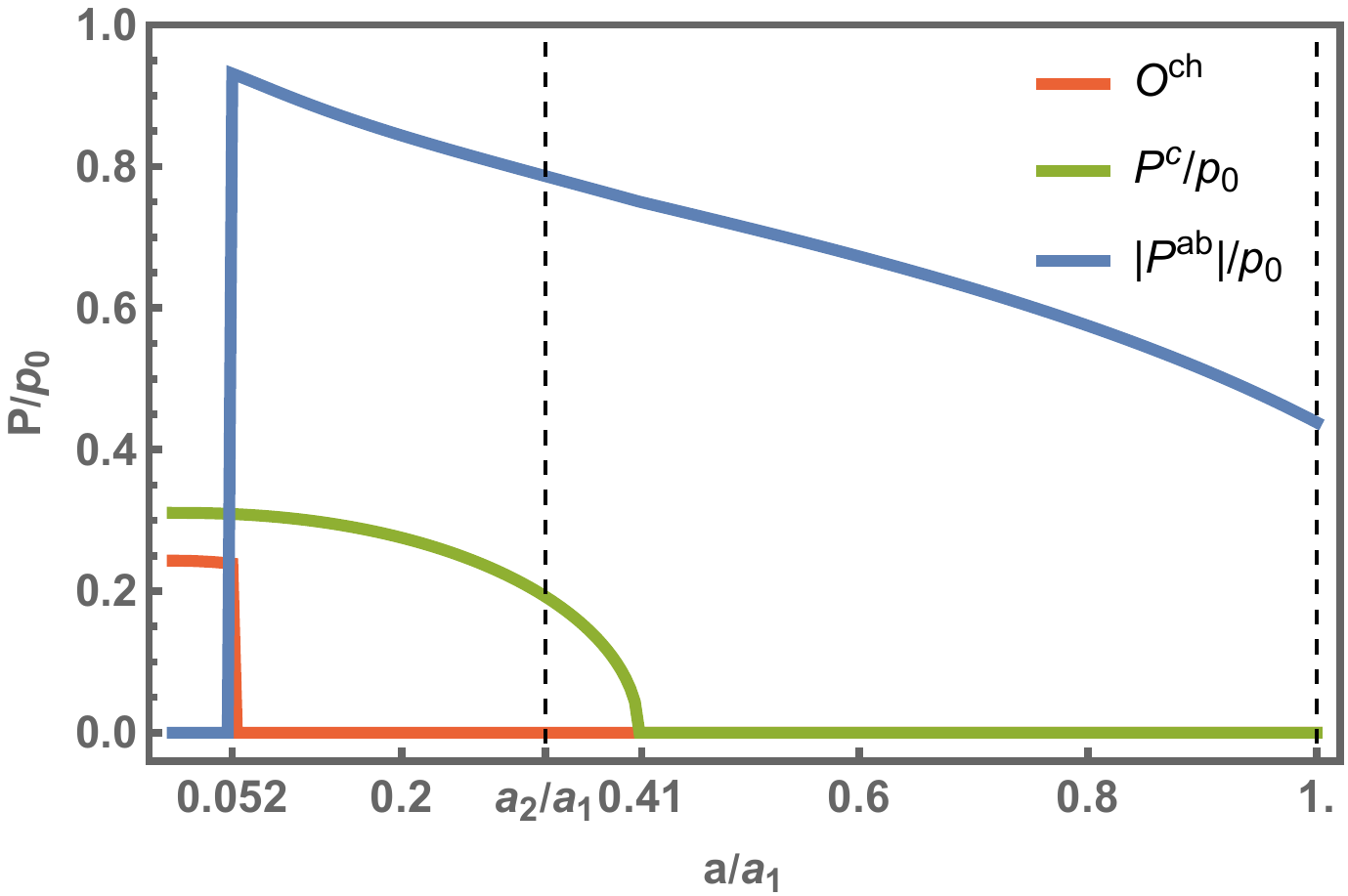}}
\end{center}
\caption{Dependence of the in-plane antiferroelectric polarization $P^{ab}$ (blue curve), the vertical polarization $P^c$ (green curve), and the chiral order parameter $O^{ch}$ (red curve) on the relative strength of quantum tunneling defined by the hopping parameter $a/a_1$ at zero temperature. The value $a/a_1=1$ corresponds to setting "1", while $a/a_{1}=0.33$ corresponds to setting "2", representing different strengths of quantum tunneling. Chiral order is suppressed for values above $a_{ch} \approx 0.052\,a_{1}$.}
\label{figaDependence}
\end{figure}

The behavior of the dipole moments at low temperatures is
strongly affected by quantum fluctuations. We included quantum fluctuations via tunneling between the local minima of the classical model.
The new parameters of the local Hamiltonian are $a$ ranging from $a_{2}\approx0.63\,\rm meV$ to $a_{1} \approx 1.9\,\rm meV$ and $d$ from
$d_{2} \approx1.26\,\rm meV$ up to $d_{1} \approx 3.9\,\rm meV$ for the hopping amplitude to the nearest minimum,
and the hopping amplitude to the next nearest minimum, respectively. The local
hopping of the unit dipole moment changes the energetic balance and
affects the response to the external electric field. We plotted the
antiferroelectric $\chi^{A}$ and ferroelectric $\chi^{Z}$ susceptibilities along
the $c$ axis in Fig.~\ref{figMFPolarizationQ}.
In the first setting $V_{b1}=56\,\rm meV$ and $\omega_1=8.2\,\rm meV$, the transition to
the planar order is shifted to a lower temperature, $T_{c1} \approx 20.2\,\rm K$. The quantum tunneling with these model parameters fully suppresses the emergence of the ferroelectric order along the $c$ axis. In contrast, the second limiting setting with $V_{b2}=176\,\rm meV$ and $\omega_2=3.9\,\rm meV$ shifts the transition temperature only to $T_{c1} \approx 30\,\rm K$ and allows for the emergence of a vertical ferroelectric order at non-zero temperature  $T_{c2} \approx 12.7\,\rm K$.
The hopping amplitudes between the local minima for the local dipole moment,
which affect the low-temperature thermodynamic behavior, were derived from the input
physical parameters  $V_{b}$ and $\omega$. The former was taken from Ref.~\cite{Kolesnikov:2016}, which describes the hexagonal symmetry of the local water dipole moment within a single beryl cavity. The amplitude connecting the two hexagons of the crystallographic cavity for the water molecule, $\omega$, was chosen to keep all the local minima equivalent. The existence of the thermodynamic order along the $c$ axis is strongly affected by the value of $\omega$, which is reflected in the hopping amplitude $a$. We plotted the dependence of the planar, vertical, and chiral order parameters on the relative hopping amplitude $a/a_1$ at zero temperature in Fig.~\ref{figaDependence}. We see that the greater the amplitude with greater quantum fluctuations, the lower the probability of the vertical order. The limiting case $a=a_{1}\approx 1.9\,\rm meV$ is beyond the edge of the existence of the vertical order at zero temperature. Therefore, we do not observe a divergence of the ferroelectric susceptibility at non-zero temperatures.
The vertical order occurs at a value of $a=0.41\, a_{1}$, and the vertical polarization is reduced to $\sim60\%$ of the original classical values at zero temperature for $a=a_{2}\approx 0.63\,\rm meV$. Chiral order at zero temperature is suppressed by quantum tunneling above $a_{ch} \approx 0.052\,a_{1}$.

%
%
%

\section{Conclusions}

We have developed a model that considers twelve equivalent orientations for the interacting dipole moments of water molecules confined in beryl crystal cavities.
Recent NMR experiments~\cite{Chlan22} revealed that the dipole
moments of type-I water molecules are not exactly orthogonal to
the hexagonal $c$ axis, but instead deflect by approximately $\pm 18^\circ$ from the crystallographic $ab$ plane.
Moreover, the potential minima for hydrogen nuclei pointing to the oxygen
atoms of the confining beryl molecules are shifted by $30^{\circ}$ in the
neighboring layers.  This situation contradicts previously used theoretical models of water molecules in beryl crystals, prompting us to introduce a model of the orientational potential for the dipole moments of water molecules with dihexagonal symmetry, involving alternating minima with the $c$-component of the dipole moment pointing up and down.

We used a classical variational mean-field approximation to obtain a qualitative thermodynamic picture of the behavior of the interacting dipole moments and their equilibrium configurations. We used the standard dipole-dipole interactions and included the lattice potential via a clock model where individual dipoles can adopt only one of the twelve minima of the lattice potential. We also used Monte Carlo simulations to more accurately include spatial fluctuations and included quantum tunneling between the minima to better assess the low-temperature macroscopic behavior.

The classical mean-field solution predicts a non-zero Curie temperature $T_{c1}$
below which the dipole moments are ferroelectrically ordered within the
crystallographic $ab$ planes, but the orientations of the planar dipole
moments are antialigned, antiferroelectrically ordered along the $c$ axis. The
thermodynamic order along this axis remains zero down to another critical point
$T_{c2}$, below which a new order emerges with a non-zero thermally averaged $c$
projection of the dipole moment. The central and most important new feature of our model, hitherto absent in existing models with only planar dipole moments in the crystallographic $ab$ plane, is the possibility of the existence of a non-zero $c$ projection of the thermally averaged dipole moment on the $c$ axis. The order along the $c$ axis leads to unexpected consequences for the symmetry of the equilibrium state. The ferroelectric order, together with the non-isotropic character of the dipole-dipole interaction, produces a helical structure of the dipole moments in which their directions go through all the local minima along the $c$ axis within the unit cell containing $n_{S}=12$ positions symmetrically distributed to cover the whole angle of $360\degree$.


Monte Carlo simulations, treating spatial fluctuations more realistically, essentially supported the thermodynamic picture of the water dipole moments obtained from classical variational mean-field theory. However, the simulations suggested a less straightforward transition to the ferroelectrically ordered $c$ projection of the polarization.
An intermediate state with a partial order of ferroelectric vertical chains was predicted, smearing the transition to the bulk ferroelectric state with a non-zero $c$-projection of the polarization and making it of the first-order type. More precise simulations of spatial fluctuations are necessary to clarify this issue.

The classical model was improved by quantum tunneling to better assess the low-temperature asymptotics. The long-range order along the $c$-axis appears to be significantly breached by quantum fluctuations that ultimately destroy the ordering at nonzero temperatures for the limiting values of the parameters in the setting "1".
However, accurately assessing the impact of quantum fluctuations is currently difficult due to the lack of experimental data from which we could extract the necessary microscopic model parameters for the hopping amplitudes $a$ and $d$ between the local minima of the dipole moments of the confined water molecules in beryl. Additionally, a direct comparison with the experiment is hindered by the fact that the water molecules fill only some of the cavities of the beryl crystals, which is not accounted for in our model. Hence, the experimental critical temperatures are overestimated in our theory depending on the hydration level. Nevertheless, our study clearly shows that the observed deflection of the dipole moments from the crystallographic $ab$-plane may lead to a nontrivial structure with a non-zero polarization along the $c$-axis, which is worth further experimental exploration.

\section*{Acknowledgments}
This work was supported by the Czech Science Foundation (project No.~20-1527S).

\end{document}